\RequirePackage{fix-cm}
\documentclass[twocolumn,epjc3, final]{svjour3}  
\smartqed  
\RequirePackage{graphicx}
\RequirePackage{amsmath}
\RequirePackage{amssymb}
\RequirePackage{multirow}
\journalname{Eur. Phys. J. C}
\begin{document}

\title{Numerical estimate of minimal active-sterile neutrino mixing for sterile neutrinos at  GeV scale}


\author{Igor Krasnov\thanksref{e1, addr1}
        \and
        Timofey Grigorin-Ryabov\thanksref{e2, addr2} 
}

\thankstext{e1}{e-mail: iv.krasnov@physics.msu.ru}
\thankstext{e2}{e-mail: timagr615@gmail.com}


\institute{Institute for Nuclear Research of the Russian Academy of Sciences, 60th October Anniversary Prospect 7a, 117312, Moscow, Russia \label{addr1}
           \and
          Department of Particle Physics and Cosmology, Physics Faculty, Moscow State University, Vorobjevy Gory 1-2, 119991, Moscow, Russia \label{addr2}
}

\date{Received: date / Accepted: date}

\maketitle

\begin{abstract}
Seesaw \,mechanism\, constrains\, from\, below\\ mixing between active and sterile neutrinos for fixed sterile neutrino masses.
Signal events associated with sterile neutrino decays inside a detector at fixed target experiment are suppressed by the mixing angle to the power of four.
Therefore sensitivity of experiments such as SHiP and DUNE should take into account minimal possible values of the mixing angles.
We extend the previous study of this subject \cite{Gorbunov:2013dta} to a more general case of non-zero CP-violating phases in the neutrino sector.
Namely, we provide numerical estimate of minimal value of mixing angles between active neutrinos and two sterile neutrinos with the third sterile neutrino playing no noticeable role in the mixing.
Thus we obtain a sensitivity needed to fully explore the seesaw type I mechanism for sterile neutrinos with masses below 2 GeV, and one undetectable sterile neutrino that is relevant for the fixed-target experiments.
Remarkably, we observe a strong dependence of this result on the lightest active neutrino mass and the neutrino mass hierarchy, not only on the values of CP-violating phases themselves.
All these effects sum up to push the limit of experimental confirmation of sterile-active neutrino mixing by several orders of magnitude below the results of \cite{Gorbunov:2013dta} from $10^{-10}$ -- $10^{-11}$ down to $10^{-12}$ and even to $10^{-20}$ in parts of parameter space; non-zero CP-violating phases are responsible for that.
\end{abstract}

\section{Introduction}

Neutrino oscillations clearly call for an extension of the Standard Model (SM) of particle physics.
Sterile neutrino models can provide a simple theoretical framework explaining this phenomenon, which makes them popular among many candidates for physics beyond SM. 
In this framework one commonly introduces three Majorana fermions $N_I, I= 1, 2, 3$,  sterile with respect to SM gauge interactions $SU(3)_c \times SU(2)_W \times U(1)_Y$.

One can write down the most general renormalizable sterile neutrino Lagrangian as:
\begin{equation}
\label{eq1}
\mathcal{L}=\mathnormal{i} \bar{N}_I\gamma^\mu \partial_\mu N_I-\Big(\frac{1}{2} M_I \bar{N}^c_I N_I + Y_{\alpha I}\bar{L}_\alpha \tilde{H} N_I +h.c.\Big),
\end{equation}
where $M_I$ are the Majorana masses, and $Y_{\alpha I}$ stand for the Yukawa couplings with lepton doublets $L_\alpha, \alpha= e, \mu, \tau$ and SM Higgs doublet ($\tilde{H}_a=\epsilon_{a b} H^*_b$).

When\, the\, Higgs\, field\, gains\, vacuum\, expectation\, value $\mathnormal{v}=246$ GeV, the Yukawa couplings in (\ref{eq1}) yield mixing between sterile $N_I$ and active $\nu_\alpha$ neutrino states. 
Diagonalization of the neutral fermion mass matrix provides active neutrinos with masses $m_i$ and mixing which are responsible for neutrino oscillation phenomena.
If active-sterile mixing angles are small, then active neutrino masses are double suppressed, $m \sim U^2 M$.
This is the standard seesaw type I mechanism, for more details one can address \cite{Mohapatra:1979ia}.

The seesaw mechanism implies non-zero mixing between sterile and active neutrinos.
But sterile neutrino mass scale is not fixed by this mechanism.
If sufficiently light, sterile neutrinos can be produced in weak processes and directly studied in particle physics experiments, for recent results see Troitsk $\nu$Mass \cite{Abdurashitov:2015jha}, OKA \cite{Sadovsky:2017qsr}, LHCb \cite{Antusch:2017hhu,Canetti:2014dka}, Belle \cite{Canetti:2014dka}, E949 \cite{Artamonov:2014urb}, NA62 \cite{CortinaGil:2017mqf}.

The authors of Ref. \cite{Gninenko:2013tk} suggest that in upcoming particle physics experiments sterile neutrinos of GeV scale may appear in heavy hadron decays and can be detected as they decay into light SM particles.
In proposed fixed target experiments such as SHiP \cite{Alekhin:2015byh} or DUNE \cite{Acciarri:2015uup} the main source of sterile neutrinos are D-mesons decays, so mostly only sterile neutrinos with masses below 2 GeV are produced.
We note that in a part of parameter space such sterile neutrinos can be responsible for leptogenesis in the early Universe which allows for direct laboratory tests of early time cosmology \cite{Gorbunov:2007ak,Drewes:2012ma,Hambye:2016sby}.

The minimal values of mixing between sterile and active neutrinos, consistent with the type I seesaw mechanism, have been estimated in \cite{Gorbunov:2013dta} for the case when some of sterile neutrinos are lighter than 2 GeV and CP-violating phases are set to zero.
Our estimates (and that of Ref. \cite{Gorbunov:2013dta} as well) are performed for mixing with only two sterile neutrinos: the third sterile neutrino is considered to be unobservable in the discussed experiments.
One scenario is that the third sterile neutrino can be too heavy to be produced in the mentioned experiments.
It can also be too light to be kinematically recognizable there.
Or it can be of interesting mass range, but very feebly interacting, i.e. practically decoupled.
Although we don't specifically restrict this mixing in such a way, it would include also the special case where this sterile neutrino can play a role of warm dark matter\footnote{We call that case ``warm'' dark matter as opposed to ``cold'' dark matter (as in WIMPs case) due to the fact that keV scale sterile neutrinos have significantly non-zero velocities at the equality epoch. Cosmic structure formation gives an upper limit on those velocities, but at the level of $10^{-4}$.} \cite{Adhikari:2016bei}, from cosmological constraints its mass is restricted to be in keV range.
Another widely studied special case is when the two heavier sterile neutrinos are strongly degenerate in mass: it is shown \cite{Boyarsky:2009ix} that such sterile neutrinos can be used to provide explanation for leptogenesis (this model is called neutrino minimal extension of the SM or just $\nu$MSM).
There are other sterile neutrino models that explain leptogenesis, and, consequently, baryon asymmetry of the Universe, for example, through Higgs doublet decay \cite{Hambye:2016sby}.
In this paper we don't introduce any special restrictions, but for numerical analysis we choose $M_1=500$ MeV, $M_2=600$ MeV, and then study how the results change with masses.

For the sterile neutrino masses about 500 MeV current upper limit on active-sterile mixing angles come from CHARM experiment \cite{Bergsma:1985is} at $|U_e|^2, |U_\mu|^2 \sim 10^{-6}$.
It can be improved in the near future by SHiP experiment \cite{Alekhin:2015byh}, which plan is to reach $|U_e|^2, |U_\mu|^2 \sim 10^{-9}$.
In more detail present bounds and expected sensitivities of some future experiments one can found, for example, in Ref. \cite{Alekhin:2015byh}.

The lower limit on mixing is usually associated with seesaw mechanism, Big Bang Nucleosynthesis (BBN) and some limits that are specific to concrete model. 
The seesaw limit is a plain mathematical limitations imbued on model \eqref{eq1} to make it consistent with current central values for active neutrino parameters.
This limit is the primary object of this study.
Independent limit comes from cosmological role of sterile neutrinos.
If at least one sterile neutrino decays during BBN, products of decay would change light element abundances.
Observation of these abundances gives us limit on contribution of non-standard BBN scenarios.

The goal of this paper is to study the dependence of minimal mixing, consistent with the seesaw mechanism, on CP-violating phases, unaccounted before in Ref. \cite{Gorbunov:2013dta}.
During this research it also became obvious that the role of lightest active neutrino mass can be essential. 
The obtained results can be used to estimate the sensitivity of future experiments required to fully explore the parameter space of type I seesaw models with sterile neutrinos in the interesting mass range.

\section{Groundwork for calculation}
\label{groundwork}

\subsection{Active sector}
\label{active}
Before going into details of concrete sterile neutrino model we provide the parametrization of active neutrino sector used in this paper.

For active neutrino masses, we use diagonal matrix $m_\nu \equiv diag\{m_1, m_2, m_3 \}$.
Experiments provide us with two related parameters and $3\sigma$ allowed ranges (see p. 248 of \cite{Patrignani:2016xqp}):
\begin{equation}
\label{data2}
\begin{array}{cccc}
\Delta m_{21}^2 \big[ 10^{-5} \, \textrm{eV}^2\big] & = &  7.37, & 6.93 - 7.97\\
|\Delta m^2| \big[ 10^{-3} \, \textrm{eV}^2\big] & = &  2.50, & 2.37 - 2.63\\
&& (2.46), & (2.33 - 2.60)
\end{array}
\end{equation}
Basically, $m_2^2 - m_1^2 = \Delta m_{21}^2$ and $m_3^2 - \frac{m_1^2 + m_2^2}{2} = \Delta m^2$.
It is usually defined that $m_2>m_1$ (just for convenience), but we don't know which of $m_1, m_3$ is smaller.
Hence, the sign of $\Delta m^2$ is unknown, and two different cases have to be considered: the \emph{normal hierarchy} case $m_1<m_2<m_3$ and the \emph{inverted hierarchy} case $m_3<m_1<m_2$.
Hereafter the values (values in brackets) correspond to normal (inverted) hierarchy of the active neutrino masses.
If there is no difference, we don't use brackets at all.
Absolute value of the lightest mass $m_{lightest}$ differs greatly from model to model and is not specified by present experiments.
So in this paper it is treated as one of the free parameters.

For the normal hierarchy we have:
\begin{displaymath}
\begin{array}{rcl}
m_1 & = & m_{lightest}\\
m_2 & = & \sqrt{m_{lightest}^2 + \Delta m_{21}^2}\\
m_3 & = & \sqrt{m_{lightest}^2 + \frac{1}{2} \Delta m_{21}^2  + |\Delta m^2|},
\end{array}
\end{displaymath}
and for the inverted hierarchy:
\begin{displaymath}
\begin{array}{rcl}
m_3 & = & m_{lightest}\\
m_1 & = & \sqrt{m_{lightest}^2 - \frac{1}{2} \Delta m_{21}^2  + |\Delta m^2| }\\
m_2 & = & \sqrt{m_{lightest}^2 + \frac{1}{2} \Delta m_{21}^2  + |\Delta m^2| }.
\end{array}
\end{displaymath}
Cosmology constrains sum of the masses from above as \cite{Ade:2015xua}:
\begin{equation}
\label{data3}
\sum_{i=1}^3 m_i < 0.57 \, \mathrm{eV}.
\end{equation}
We choose $m_{lightest}<0.2\,$eV as the approximate constraint equivalent to (\ref{data3}).

The transformation from flavour basis to massive basis is provided by hermitian conjugate of Pontecorvo-Maki-Nakagawa-Sakata unitary mixing matrix $U_{PMNS}$:
\begin{equation}
\label{eq2}
\left( \begin{array}{ccc}
\nu_1\\
\nu_2\\
\nu_3\\
\end{array} \right)
= U_{PMNS}^\dagger \left(\begin{array}{ccc}
\nu_e\\
\nu_\mu\\
\nu_\tau\\
\end{array}\right),
\end{equation}
which in turn, can be parametrized as follows (see p. 248 of  \cite{Patrignani:2016xqp}):
\begin{equation}
\label{eq3}
\begin{array}{rcl}
U_{PMNS}^\dagger& = & \left(\begin{array}{ccc}
1 & 0 & 0\\
0 & c_{23} & s_{23}\\
0 & -s_{23} & c_{23}\\
\end{array}\right) \times \left(\begin{array}{ccc}
c_{13} & 0 & s_{13} e^{- i \delta}\\
0 & 1 & 0\\
-s_{13} e^{i \delta} & 0 & c_{13}\\
\end{array}\right)\times\\
&&\times\left(\begin{array}{ccc}
c_{12} & s_{12} & 0\\
-s_{12} & c_{12} & 0\\
0 & 0 & 1\\
\end{array}\right) \times  \left(\begin{array}{ccc}
e^{- i \frac{\alpha_1}{2}} & 0 & 0 \\
0 & e^{- i \frac{\alpha_2}{2}} & 0\\
0 & 0 & 1\\
\end{array}\right).
\end{array}
\end{equation} 
Here $c_{ij}$ and $s_{ij}$ stand for $\cos\theta_{ij}$ and $\sin\theta_{ij}$, with $i,j=1,2,3, i<j$.

The angles entering \eqref{eq3} have been experimentally determined.
We take best fit values and $3\sigma$ allowed ranges for $\sin^2 \theta_{ij}$ (see p. 248 of \cite{Patrignani:2016xqp}):
\begin{displaymath}
\begin{array}{cccc}
\sin^2 \theta_{12} & = & 0.297, & 0.250  - 0.354\\
\sin^2 \theta_{23} & = & 0.437, & 0.379 -0.616\\
&& (0.569), & (0.383 -0.637)\\
\sin^2 \theta_{13} & = & 0.0214, & 0.0185 - 0.0246\\
&& (0.0218), & (0.0186 -0.0248). 
\end{array}
\end{displaymath}
As it is more convenient to use angles $\theta_{ij}$ themselves, rather than the values of $\sin^2 \theta_{ij}$, we list them here (corresponding to the best fit value):
\begin{equation}
\label{data1}
\begin{array}{cccc}
\theta_{12} & = & 33.02^\circ & \\
\theta_{23} & = & 41.38^\circ & (48.97^\circ)\\
\theta_{13} & = & 8.41^\circ & (8.49^\circ)
\end{array}
\end{equation}

CP-violating phases $\delta, \alpha_1, \alpha_2$ entering (\ref{eq3}) are still not specified by experiments as strictly as angles $\theta_{ij}$ and are one of the main subjects of study in this paper.
In the most general case we have $\delta \in [0,2 \pi), \alpha_1 \in [- \pi, \pi), \alpha_2 \in [- \pi, \pi)$.
For the Dirac phase $\delta$ we have the best fit value (see p. 248 of \cite{Patrignani:2016xqp}):
\begin{equation}
\label{data4}
\begin{array}{cccc}
\delta / \pi & = & 1.35  &  (1.32)
\end{array}
\end{equation}
At $3\sigma$ no physical values of $\delta$ are disfavoured (see p. 248 of \cite{Patrignani:2016xqp}).
Basically we treat $\delta$ as a free parameter, but always provide graph for best fit value (\ref{data4}) if possible.
Majorana phases $\alpha_1, \alpha_2$ haven't yet been observed in any experiment.
Consequently, they are considered as free parameters, for more detail see Sec. \ref{numerical}.

\subsection{Sterile sector}
\label{sterile}
Next we move on to the subject of sterile neutrino sector parametrization.

It is convenient to adopt the bottom-up parametrization for the $3 \times 3$ Yukawa coupling
matrix $Y$ \cite{Casas:2001sr}:

\begin{equation}
\label{eq4}
Y \equiv \frac{\mathnormal{i}\sqrt{2}}{\mathnormal{v}} M_R^{\frac{1}{2}} R m_\nu^{\frac{1}{2}} U_{PMNS}^\dagger,
\end{equation}
where $M_R \equiv diag\{M_1, M_2, M_3 \}$.

Sterile neutrino mass scale is not fixed in the seesaw mechanism, and in this paper we use for the numerical simulations values from the mass range $M_I< 2\,$ GeV, most relevant for the upcoming fixed target experiments.
We discuss different cases of sterile neutrino mass spectrum in Sec. \ref{matrix}.
Matrix $R$ is a complex orthogonal matrix, $R^T R =1_3$.
We use the following parametrization of $R$:
\begin{equation}
\label{eq5}
\begin{array}{rcl}
R & = &  diag\{\pm 1, \pm 1, \pm 1\} \times \left( \begin{array}{ccc}
1 & 0 & 0\\
0 & c_3 & s_3\\
0 & -s_3 & c_3\\
\end{array}\right)\times\\
&& \times \left(\begin{array}{ccc}
c_2 & 0 & s_2\\
0 & 1 & 0\\
-s_2 & 0 & c_2\\
\end{array}\right)\times\left(\begin{array}{ccc}
c_1 & s_1 & 0\\
-s_1 & c_1 & 0\\
0 & 0 & 1\\
\end{array}\right),
\end{array}
\end{equation}
where $c_i =\cos z_i, s_i =\sin z_i$, and $z_i \subset \mathbb{C}$ are not restricted in any way.

\section{Matrix of mixing angles}
\label{matrix}
The main subject of this work is the matrix of mixing angles between active and sterile neutrinos:
\begin{equation}
\label{eq6}
U= \frac{\mathnormal{v}}{\sqrt{2}}M_R^{-1} Y = \mathnormal{i}M_R^{-\frac{1}{2}} R m_\nu^{\frac{1}{2}} U_{PMNS}^\dagger
\end{equation}
It depends on three complex angles $z_i$ entering (\ref{eq5}) and three yet unknown CP-violating phases of $U_{PMNS}$ matrix (\ref{eq3}).
Also we know to a certain extent three $U_{PMNS}$ matrix angles (\ref{data1}) and two differences in active neutrino masses squared (\ref{data2}), but neither mass of the lightest neutrino $m_{lightest}$ nor the hierarchy of masses.

We consider sterile neutrino production in a fixed-target experiment due to mixing (\ref{eq6}) in weak decays of hadrons. 
As D-meson decays are the main source of sterile neutrinos in the mentioned fixed target experiments, sterile neutrinos with $M_I> 2\,$ GeV are too heavy to be produced.
Sterile neutrino main signature is a weak decay into SM particles due to the same mixing.
The number of signal events depends on the values of $| U_{I\alpha} |^2, I=1,2,3; \alpha= e, \mu, \tau$.
Obviously, the sign matrix in (\ref{eq5}) can be omitted during calculation of $| U_{I\alpha} |^2$. 
We should note that due to kinematics, mixing $| U_{I\tau} |^2$ doesn't play any role in the decays\footnote{Here we count only observable decay modes; $|U_{I\tau}|^2$ governs decay into unrecognisable $\tau$-neutrino} of sterile neutrino emerged in decays of charmed hadrons.
Hence we study the minimal values of $| U_{Ie} |^2$ and $|U_{I\mu} |^2$ in order to determine what maximal sensitivity the coming experiments should achieve to fully explore the type I seesaw model.

The case of one sterile neutrino being lighter than 2 GeV (e.g. $M_1>2\,\textrm{GeV}$, $M_2>2\,\textrm{GeV}$, $M_3<2\,\textrm{GeV}$) is of little interest as sterile neutrinos $N_1, N_2$ can't be observed in the discussed experiments in this case.
During the scan of possible values of unrestricted parameters $z_1, z_2, z_3, \alpha_1, \alpha_2$, there always can be found such a set of these parameters, that $| U_{3e} |^2=0, | U_{3\mu} |^2=0$.
Because of that we can't rule out this model even if a fixed-target experiment doesn't detect sterile neutrinos with ultimately high precision.
In this paper we only consider the case when one of sterile neutrinos doesn't have significant contribution to the mixing with active sector.
In case when $M_1, M_2, M_3 \lesssim\,$2 GeV it has to be specifically implied, as all three sterile neutrinos kinematically can be produced in D-meson decays.
This noninteracting sterile neutrino can serve as a dark matter candidate, as it is decoupled from others \cite{Adhikari:2016bei}.
If DM sterile neutrino are produced by oscillation in primordial plasma, its mass scale is in keV-range \cite{Adhikari:2016bei}.
From experimental point of view such sterile neutrinos can't be kinematically recognizable in fixed-target experiments.
If one wants to stay in the boundaries of $\nu$MSM to simultaneously provide dark matter candidate and leptogenesis in the early Universe, one needs two heavy sterile neutrinos masses to be degenerate.
It was suggested, though, that leptogenesis could be successful in a much wider range of masses of sterile neutrinos, given that they are at the same scale, including GeV region, and mix to the active neutrino with comparable strength \cite{Drewes:2012ma}.
It is estimated in \cite{Canetti:2014dka} for $M_I<5\,$GeV case that mixing $U^2_{\mu I} \lesssim 10^{-10}$ is consistent with the leptogenesis scenario.
The lower limits on mixing for $M_1, M_2, M_3 <\,$2 GeV considered in \cite{Canetti:2014dka} are $|U_{\mu I}|^2 \sim 10^{-13}$ for $m_{lightest} = 0.23\,$eV and $|U_{\mu I}|^2 \sim 10^{-12}$ for $m_{lightest} = 0$. 
Study of the case of all three masses being below 2 GeV scale, $M_1, M_2, M_3 <\,$2 GeV and none of the sterile neutrinos being decoupled from the active fermions (and so potentially discoverable in a beam-dump experiments), can be a subject for further research.

Lastly, the third neutrino can be heavy $M_1<2\,$GeV, $M_2<2\,$GeV, $M_3>2\,$GeV.
Naturally, in this case $N_3$ kinematically can not be produced in D-meson decays and has no effect in these experiments.

So for our setup the relevant observables are mixing angels between $N_1, N_2$ and $\nu_e, \nu_\mu$.
Our aim is to find the lowest sensitivity enough to rule out the seesaw mechanism.
It implies the absence of any signal of either of sterile neutrinos.
Hence the relevant combinations to constrain are:
\begin{equation}
\label{eq7}
\begin{array}{ccc}
U_e \equiv |U_{1 e}|^2 + |U_{2 e}|^2\\
U_\mu \equiv |U_{1 \mu}|^2  + |U_{2 \mu}|^2\\
\end{array}
\end{equation}
Thus we search for minimal values of $U_e, U_\mu$, which, at a given $M_1, M_2$ guarantee full exploration of the seesaw mechanism for such case. 
It can be seen that $U_e$ and $U_\mu$ don't depend on $M_3$ in this particular case.
In our numerical studies, unless stated otherwise, we set $M_1=500$ MeV, $M_2=600$ MeV.
We discuss what happens for other spectra in Sec. \ref{masses}.

We point out that from eq. \eqref{eq6} one can see that our results can be rescaled to other mass scales.
If one simultaneously changes sterile neutrino masses by factor $X$: $M_I \to X M_I$, than mixing also simply changes by that factor: $U_e \to \frac{1}{X} U_e, U_\mu \to \frac{1}{X} U_\mu$.
We choose mass scale that can be tested in proposed fixed target experiments \cite{Alekhin:2015byh,Acciarri:2015uup}, but our results can be simply scaled for the case of heavier sterile neutrinos.
Note that for heavier sterile neutrinos the main source of production is not the meson decays, but the decays of heavier SM particles, e.g. weak gauge bosons produced in colliders such as LHC, FCC.

\section{BBN constraint}
\label{BBN}
We should note, that for sterile neutrinos at GeV scale we have constraints from the Big Bang Nucleosynthesis.
They follow from the fact that sterile neutrino decay products would change light element abundances originating from BBN.
Sterile neutrinos can be born in the early Universe, although we don't consider any specific mechanism in this paper.
They are not stable due to mixings with active neutrino, and may decay during BBN.
SM products of sterile neutrino decays are very energetic and can destroy atoms that has already been produced, thus changing chemical composition of the Universe. 
Direct observations imply limits on how much new physics can affect these abundances.

These limitations are mainly independent from the seesaw constraint, and can significantly change with the introduction of some new physics affecting active-sterile neutrino mixing in the early Universe.
One such example is the inflation theories which introduce coupling of the inflaton to sterile neutrinos, such as Ref. \cite{Shaposhnikov:2006xi}.

One can consider two realistic scenarios with small mixing.
The first is that mixing is significant enough for sterile neutrinos to come to equilibrium and depart from it in the early Universe before BBN, and then, by the time of BBN, decay in SM particles.
In this case we can obtain lower limit on the mixing.
Second scenario is when mixing is greatly suppressed and sterile neutrinos never equilibrate.
In such a case if mixing with active sector is lowered even further, BBN can no longer restrict mixing in this area.

First of all, the production rate of sterile neutrinos can be expressed as \cite{Barbieri:1989ti,Dolgov:2000pj}:
\begin{equation}
\label{eqBBN1}
\Gamma \sim G_F^2 \frac{\sin^2 2\theta}{\left(1+ c(T) \times 10^{-7} \Big(\frac{T}{\textrm{GeV}}\Big)^6 \Big(\frac{M_I}{\textrm{GeV}}\Big)^{-2}\right)^2}T^5,
\end{equation}
where $T$ is plasma temperature, $G_F$ is the Fermi constant, $c(T) \sim 1$ is a numerical parameter varying slightly with temperature and $\sin^2 2\theta$ is mixing parameter for the case where only one sterile neutrino mixes with only one active neutrino.
We neglect actual numerical coefficients, differing for mixing with different active neutrinos at different temperatures, because that is of little importance for our estimate.
From this point on we use $|U|^2$ instead of $\sin^2 2\theta$ which corresponds to our case, then in total there are three sterile neutrinos.
Equilibrium is achieved at $H=\Gamma$, where $H$ is the Hubble parameter,
\begin{equation}
\label{eqBBN2}
\begin{array}{ccc}
H=\frac{T^2}{M_{Pl}^*}\\
M_{Pl}^*=\sqrt{\frac{90}{8 \pi^3 g_*}}M_{Pl} \simeq \frac{M_{Pl}}{1.66\,\sqrt{g_*}},
\end{array}
\end{equation}
with $M_{Pl}$ being the Planck mass and $g_*$ standing for the effective number of degrees of freedom in plasma.

One can obtain numerically that for $M_I=500\,$MeV the equilibrium can be achieved for $|U|^2 \gtrsim |U_b|^2 \approx 2 \times 10^{-11}$.
The boundary value $|U_b|^2$ corresponds to the situation when sterile neutrinos come into thermal equilibrium and exit it immediately at temperature $T_{eq} \sim 10\,$GeV; with smaller mixing the sterile neutrinos would never be in equilibrium.
One can get values of $|U_b|^2, T_{eq}$ by equating \eqref{eqBBN1} and \eqref{eqBBN2}, expressing $|U|^2$ as function of $T$ and finding its minima and $T$ corresponding to it. We take $c(T) =0.76$ in accordance with \cite{Dolgov:2000pj}, $g_*(10\,\textrm{GeV})=86\frac{1}{4}$.
Physically $T_{eq}$ is the temperature of maximal production in \eqref{eqBBN1}.

After decoupling, the sterile neutrino concentration is:
\begin{equation}
\label{eqBBN4}
n_I^{eq}=\frac{3}{4} 2 \frac{\zeta(3)}{\pi^2 }T^3.
\end{equation}

The smallest mixing we obtain for the seesaw model (see Figs. \ref{fig7} -- \ref{fig14}) are typically smaller than $|U_b|^2$.
The relevant case is then the second scenario: the long living sterile neutrino that never was in equilibrium.
The BBN constraints one can obtain from Ref. \cite{Kawasaki:2017bqm}.
Naturally, if sterile neutrino is never abundant enough for equilibrium, it's concentration is less than that in equation \eqref{eqBBN4}, and since $n_I \propto \Gamma \propto |U|^2$ it can be estimated as:
\begin{equation}
\label{eqBBN6}
n_I = n_I^{eq} \frac{|U|^2}{|U_b|^2}.
\end{equation}
This rough estimate is enough for our purposes.

In \cite{Kawasaki:2017bqm} the BBN constraints for decays of new particle X are introduced for the variable:
\begin{equation}
\label{eqBBN7}
\zeta_X = M_X \frac{n_X}{s},
\end{equation}
where $s=g_* \frac{4 \pi^2}{90} T^3$ is entropy density.
In our case \eqref{eqBBN7} matches to:
\begin{equation}
\label{eqBBN8}
\zeta_I = \epsilon M_I \frac{135 \zeta(3)}{8 \pi^4} \frac{g_*(1\,\textrm{MeV})}{g_*(10\,\textrm{GeV})} \frac{|U|^2}{|U_b|^2},
\end{equation}
where $g_*(10\,\textrm{GeV})=86\frac{1}{4}, g_*(1\,\textrm{MeV})= 10\frac{3}{4}$ and $\epsilon$ is the part of energy coming to concrete decay channel.

At $M_I = 100\,$MeV two decay channels are available for sterile neutrinos: $N \to \nu \bar{\nu}\nu$ and $N \to \nu e^+ e^-$.
Heavier sterile neutrinos decay into muons, pions, kaons and etc.
One can take sterile neutrino decay rates from Ref. \cite{Gorbunov:2007ak}, and obtain for the sterile neutrino lifetime approximately:
\begin{equation}
\label{eqBBN10}
\tau_I \approx 20 \frac{|U_b|^2}{|U|^2} \left(\frac{500\,\mathrm{MeV}}{M_I}\right)^5 \mathrm{sec}.
\end{equation}

To use the estimate of Ref. \cite{Kawasaki:2017bqm} we find from \eqref{eqBBN8} for the reference values of model parameters $M_I=500\,\mathrm{MeV}, |U_b|^2 =2 \times 10^{-11}$:
\begin{equation}
\label{eqBBN11}
\zeta_I = 1.3 \times 10^{-2} \left(\frac{M_I}{500\,\mathrm{MeV}}\right) \frac{|U|^2}{|U_b|^2}  \textrm{GeV}
\end{equation}
From Fig. 10 in Ref. \cite{Kawasaki:2017bqm} one can see that such parameters are excluded by the observed light element abundances.
The smaller mixings seen in Figs. \ref{fig7} -- \ref{fig12a}, $|U|^2 \sim 10^{-11} - 10^{-20}$ are excluded as well.
On the other hand at $M_I=500\,\mathrm{MeV}, |U|^2 = 10^{-23}$ we have $\zeta_I = 6.5 \times 10^{-15}\,\textrm{GeV}$ and $\tau_I=3.2 \times 10^{13}$sec.
According to Fig. 10 in Ref. \cite{Kawasaki:2017bqm}, it is outside the excluded zone.
Therefore, at least all values $10^{-23} < |U|^2 < 2 \times 10^{-11}$ are excluded by BBN.

Combining \eqref{eqBBN11} with Fig. 10 in Ref. \cite{Kawasaki:2017bqm} one can see that BBN excludes only mixing $|U|^2$ above a certain, although significantly small, value.
We show in this paper that such straightforward limitation can be a stronger constraint than seesaw mechanism constraint.
BBN constraint is independent from seesaw mechanism constraint and can be avoided with introduction of the new physics.
On side note, BBN doesn't constrain too weak mixings, which we show might also be allowed by seesaw constraint.
As we discuss in Sec. \ref{zero U_e}, this scenario can't be tested in any experiments in the near future, so we don't study it in details.
We state that our numerical estimate can't recognise values of mixing below $|U|^2 \sim 10^{-20}$, that lays in the zone still excluded by BBN \eqref{eqBBN11}.
This qualitative estimate is correct for the interesting sterile neutrino mass range $100\,\textrm{MeV} < M_I < 2\,\textrm{GeV}$.

Note in passing, as we explained at the end of Sec. \ref{matrix}, if we rescale the sterile neutrino mass range from $100\,\textrm{MeV} < M_I < 2\,\textrm{GeV}$ to $1\,\textrm{GeV} < M_I < 20\,\textrm{GeV}$, we should change the values of $|U|^2$ in \eqref{eqBBN11} by the factor of $0.1$.
From the definition, $|U_b|^2$ also changes by the factor of $0.1$, while $T_{eq}$ changes by the factor of $(10)^\frac{1}{3}\approx 2.15$.
That corresponds to $\zeta_I$ changing by the factor of 10 in \eqref{eqBBN11}.
On the other hand, sterile neutrino lifetime changes by the factor of $10^{-5}$ as seen from \eqref{eqBBN10}.
To be more sincere, the lifetime takes even lesser values as at higher masses of the sterile neutrinos new decay modes become available.
From Fig. 10 in Ref. \cite{Kawasaki:2017bqm} one can see that such changes should lessen the constraints coming from BBN for sterile neutrino with greater mass.

\section{Numerical part}
\label{numerical}

\begin{figure*}
\includegraphics[height=0.45\textheight]{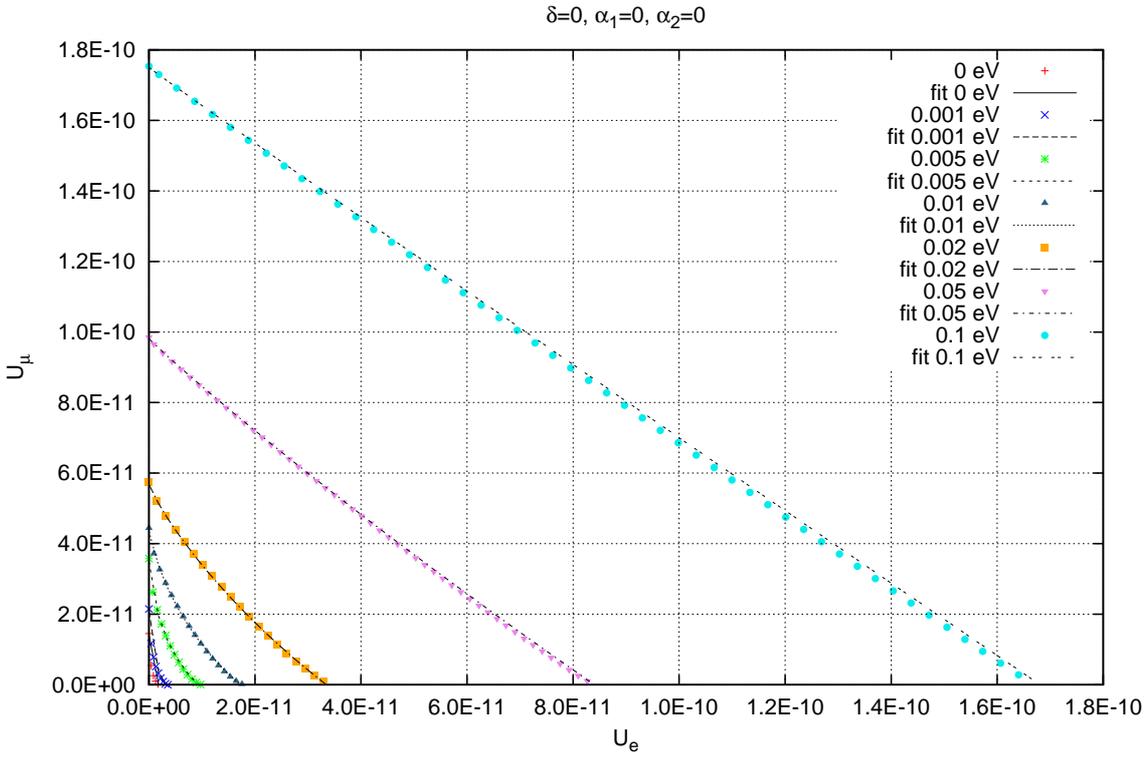}
\caption{Dependence of minimal $U_\mu$ on minimal $U_e$ for the normal hierarchy and $\delta=\alpha_1=\alpha_2=0$. Different curves correspond to different $m_{lightest}$ values.}
\end{figure*}

\begin{figure*}
\label{fig7}
\includegraphics[height=0.45\textheight]{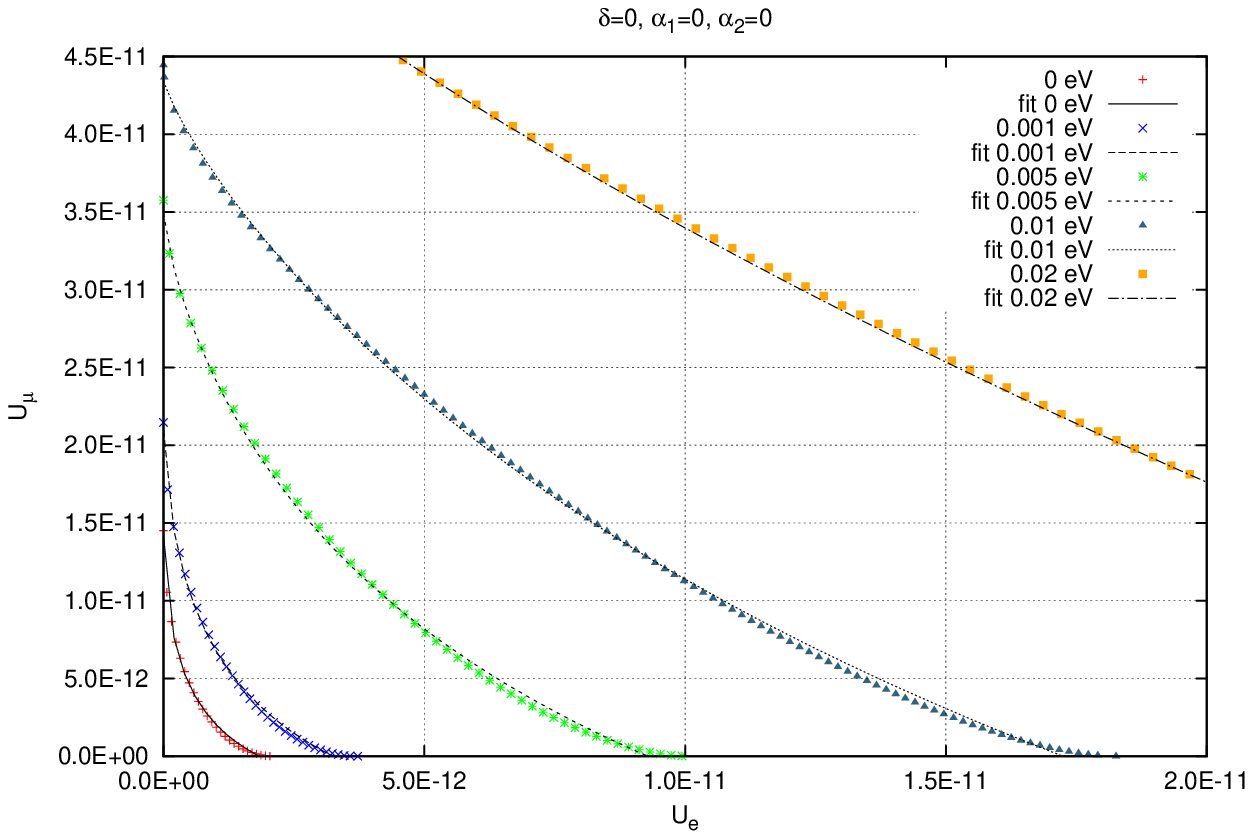}
\caption{Zoom in the small scale area of Fig \ref{fig7}.}
\label{fig7a}
\end{figure*}

\begin{figure*}
\includegraphics[height=0.45\textheight]{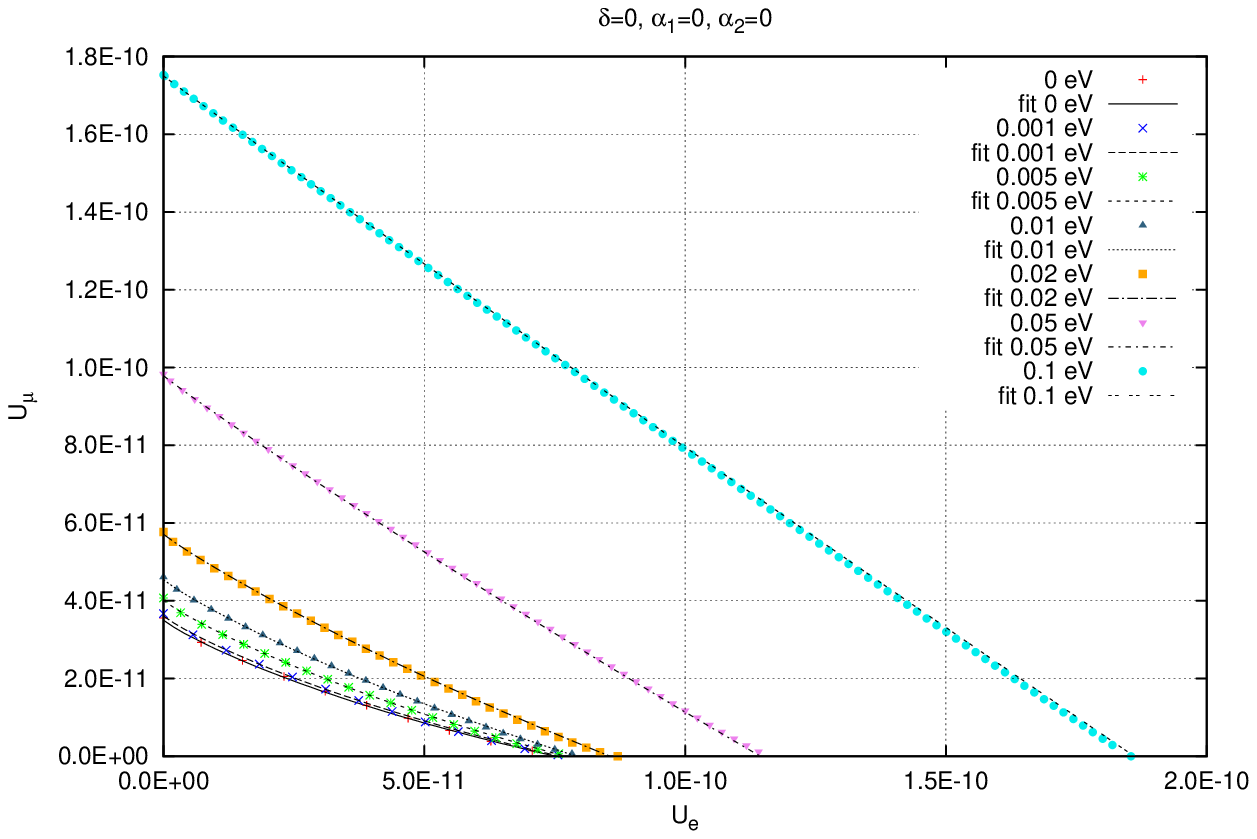}
\caption{Dependence of minimal $U_\mu$ on minimal $U_e$ for the inverted hierarchy and $\delta=\alpha_1=\alpha_2=0$. Different curves correspond to different $m_{lightest}$ values.}
\label{fig1}
\end{figure*}

\begin{figure*}
\includegraphics[height=0.45\textheight]{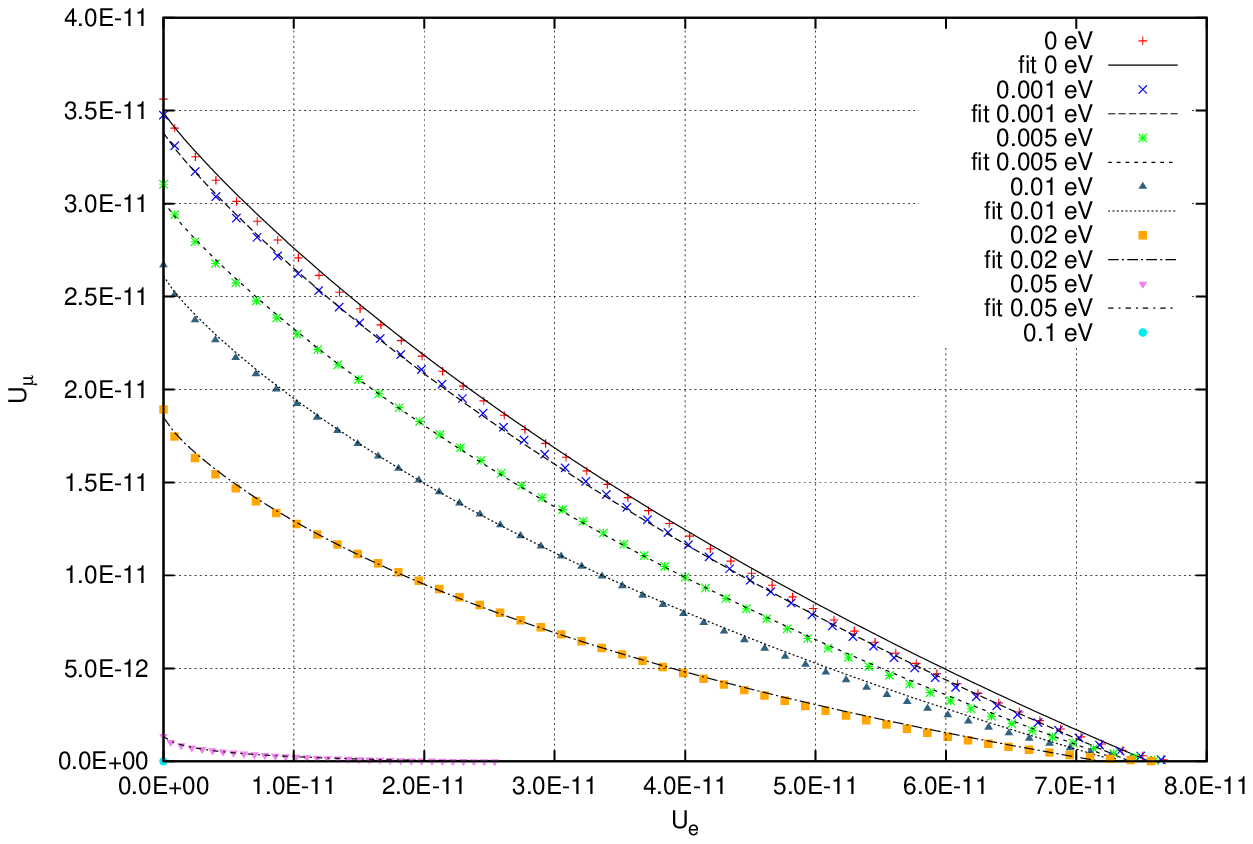}
\caption{Dependence of minimal $U_\mu$ on minimal $U_e$ for the inverted hierarchy. Different curves correspond to different $m_{lightest}$ values, $\delta, \alpha_1, \alpha_2$ are minimization variables.}
\label{fig6}
\end{figure*}

\begin{figure*}
\includegraphics[height=0.45\textheight]{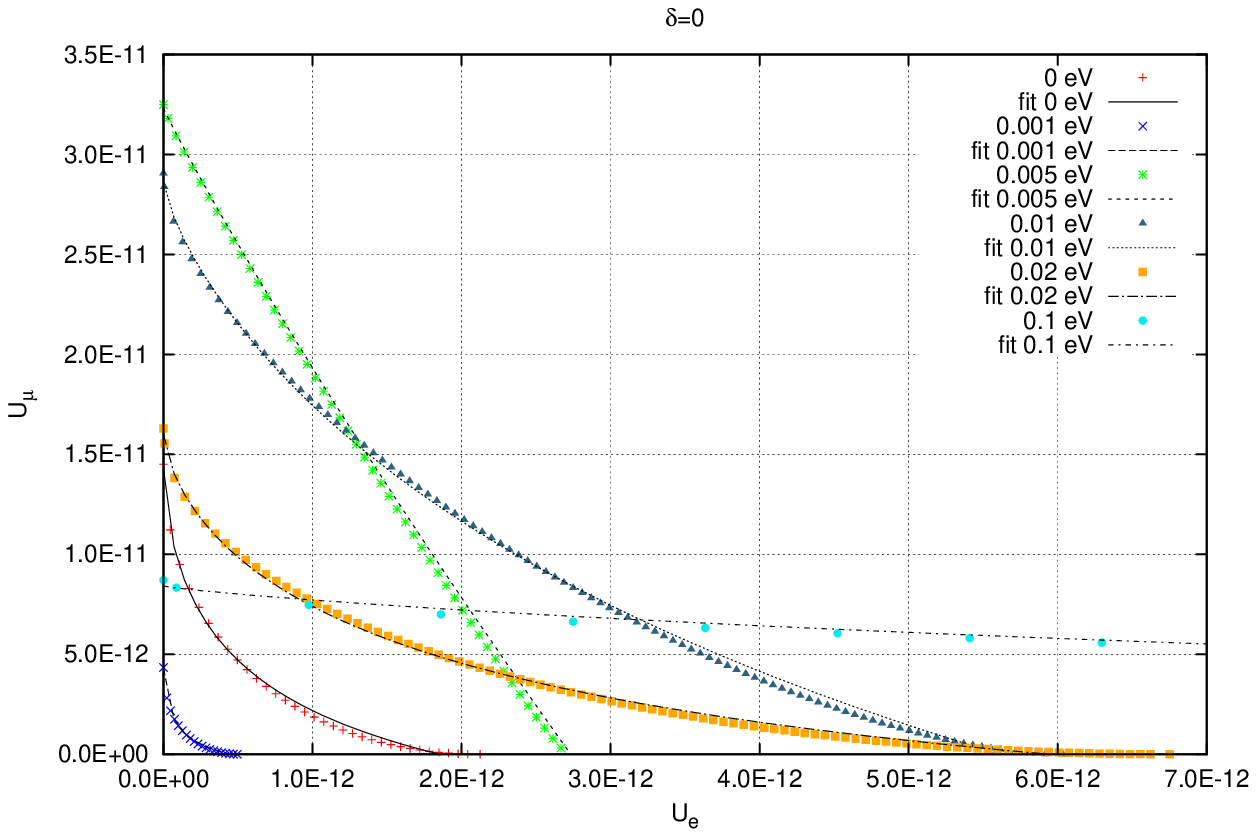}
\caption{Dependence of minimal $U_\mu$ on minimal $U_e$ for the normal hierarchy and $\delta=0$. Different curves correspond to different $m_{lightest}$ values, $\alpha_1, \alpha_2$ are minimization variables.}
\label{fig8a}
\end{figure*}

\begin{figure*}
\includegraphics[height=0.45\textheight]{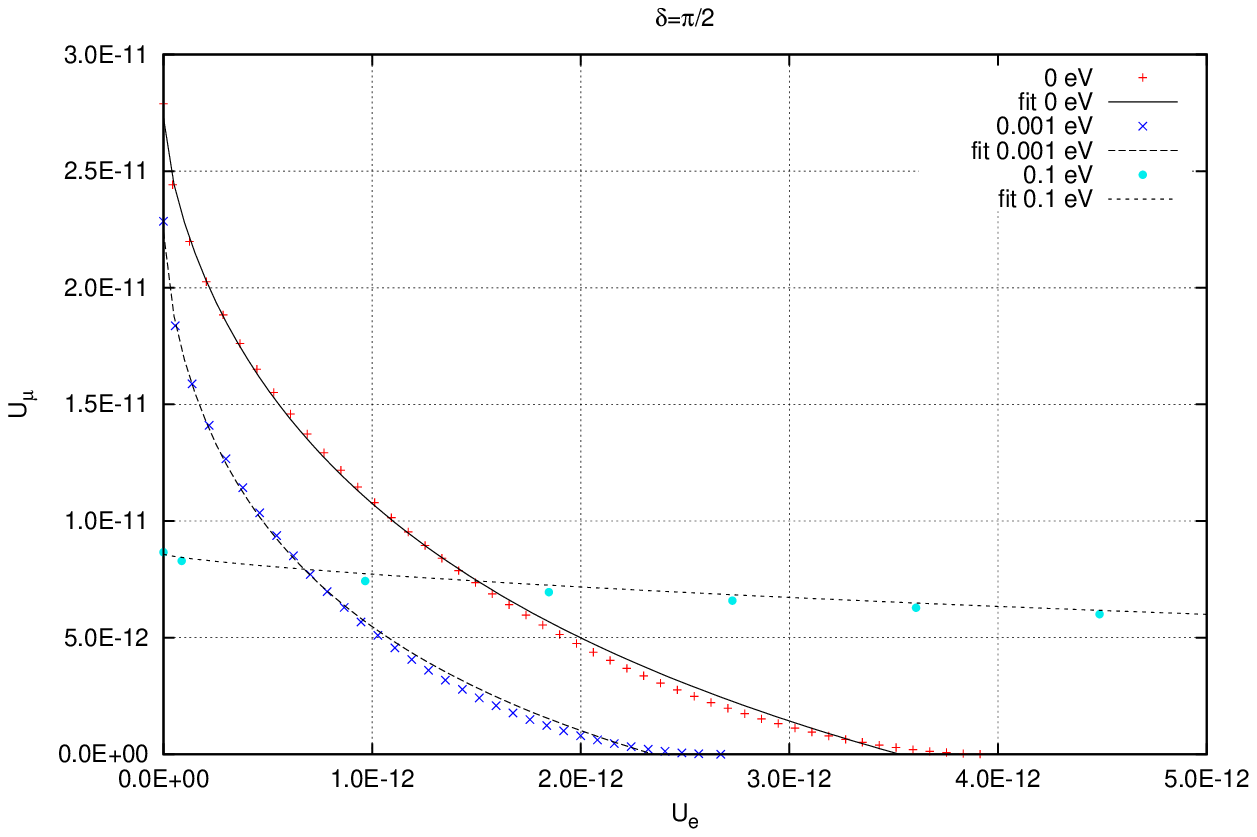}
\caption{Dependence of minimal $U_\mu$ on minimal $U_e$ for the normal hierarchy and $\delta=\frac{\pi}{2}$. Different curves correspond to different $m_{lightest}$ values, $\alpha_1, \alpha_2$ are minimization variables.}
\label{fig9a}
\end{figure*}

\begin{figure*}
\includegraphics[height=0.45\textheight]{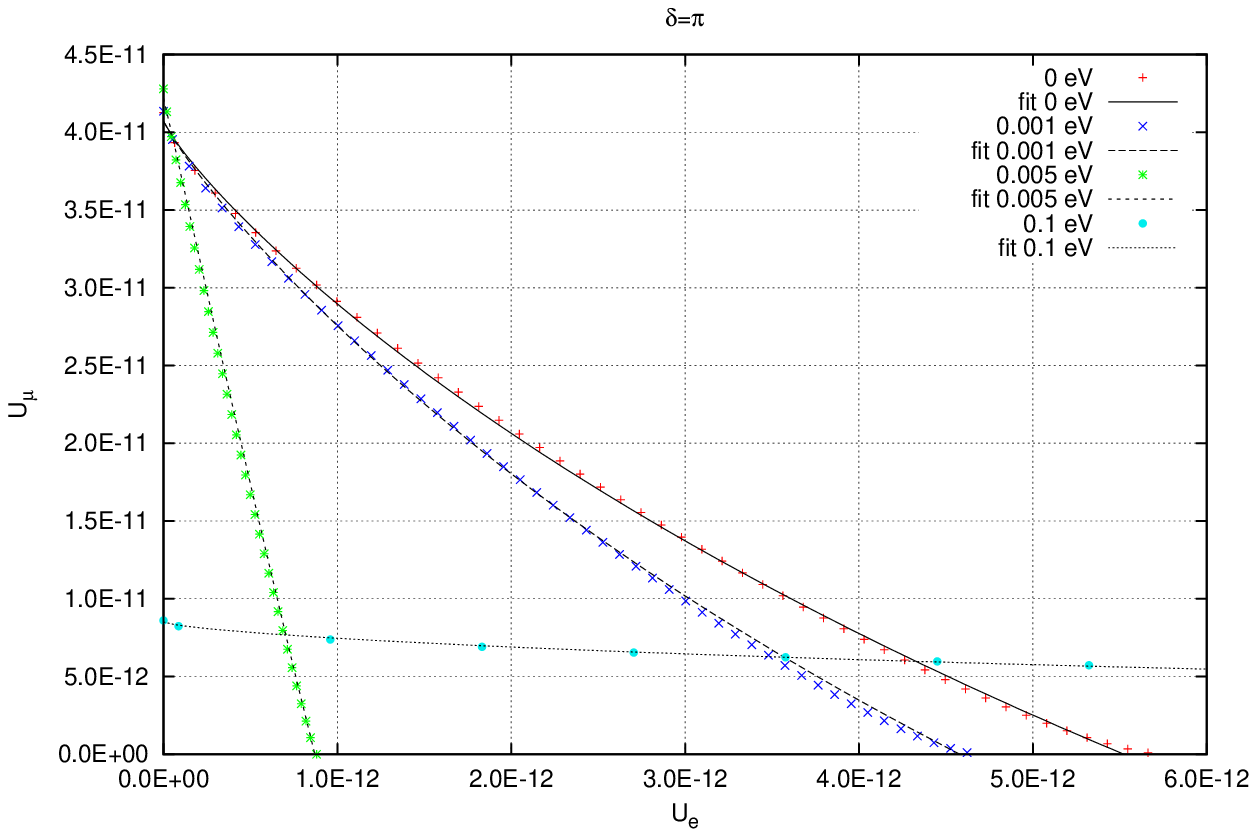}
\caption{Dependence of minimal $U_\mu$ on minimal $U_e$ for the normal hierarchy and $\delta=\pi$. Different curves correspond to different $m_{lightest}$ values, $\alpha_1, \alpha_2$ are minimization variables.}
\label{fig10a}
\end{figure*}

\begin{figure*}
\includegraphics[height=0.45\textheight]{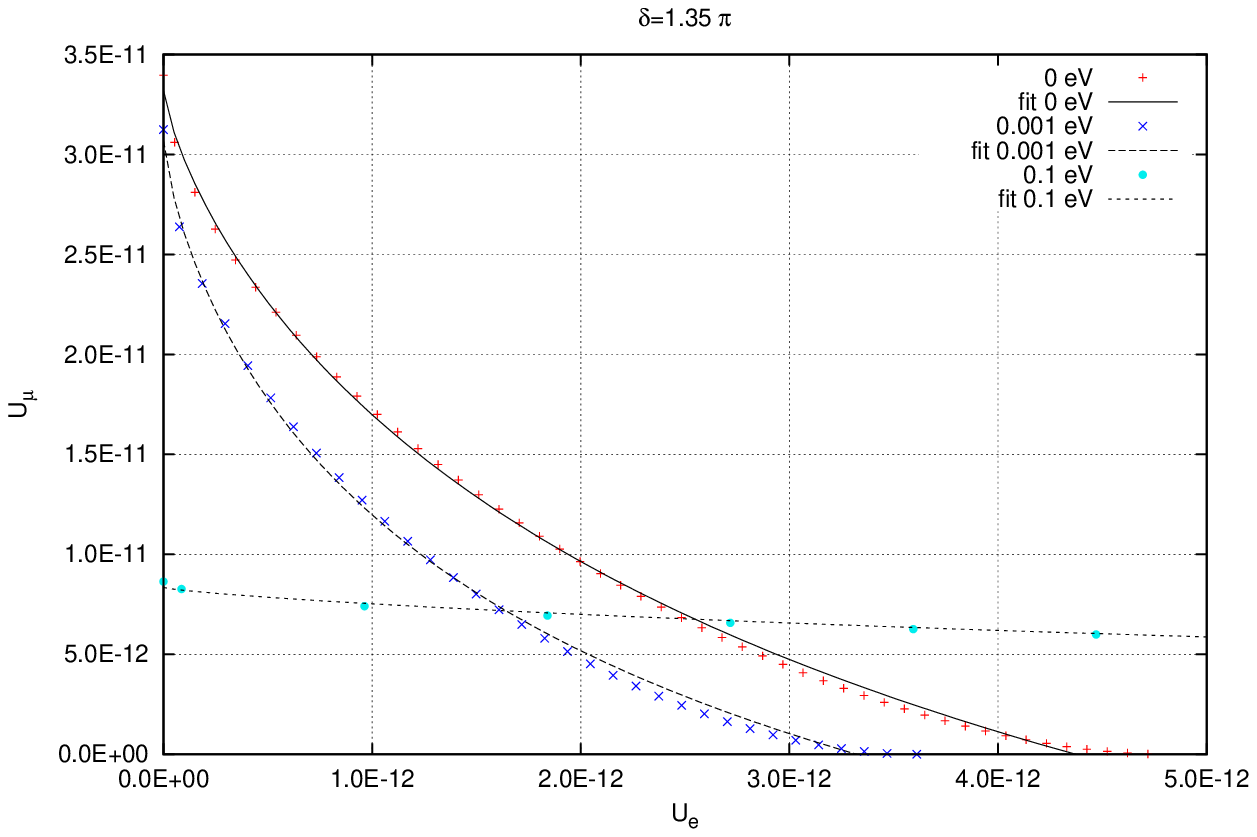}
\caption{Dependence of minimal $U_\mu$ on minimal $U_e$ for the normal hierarchy and $\delta=1.35 \pi$ (the best fit value for the normal hierarchy). Different curves correspond to different $m_{lightest}$ values, $\alpha_1, \alpha_2$ are minimization variables.}
\label{fig11a}
\end{figure*}

\begin{figure*}
\includegraphics[height=0.45\textheight]{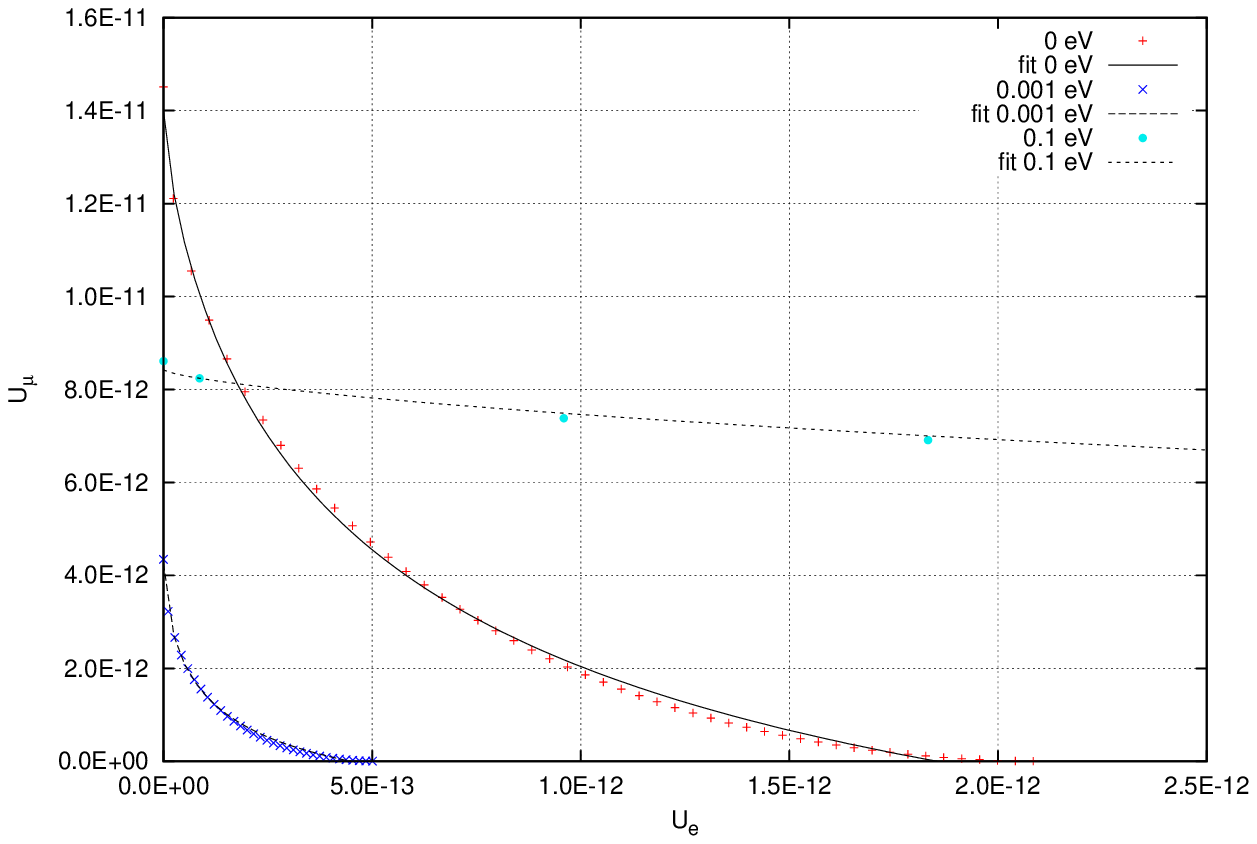}
\caption{Dependence of minimal $U_\mu$ on minimal $U_e$ for the normal hierarchy. Different curves correspond to different $m_{lightest}$ values, $\delta, \alpha_1, \alpha_2$ are minimization variables.}
\label{fig12a}
\end{figure*}

\begin{figure*}
\includegraphics[height=0.45\textheight]{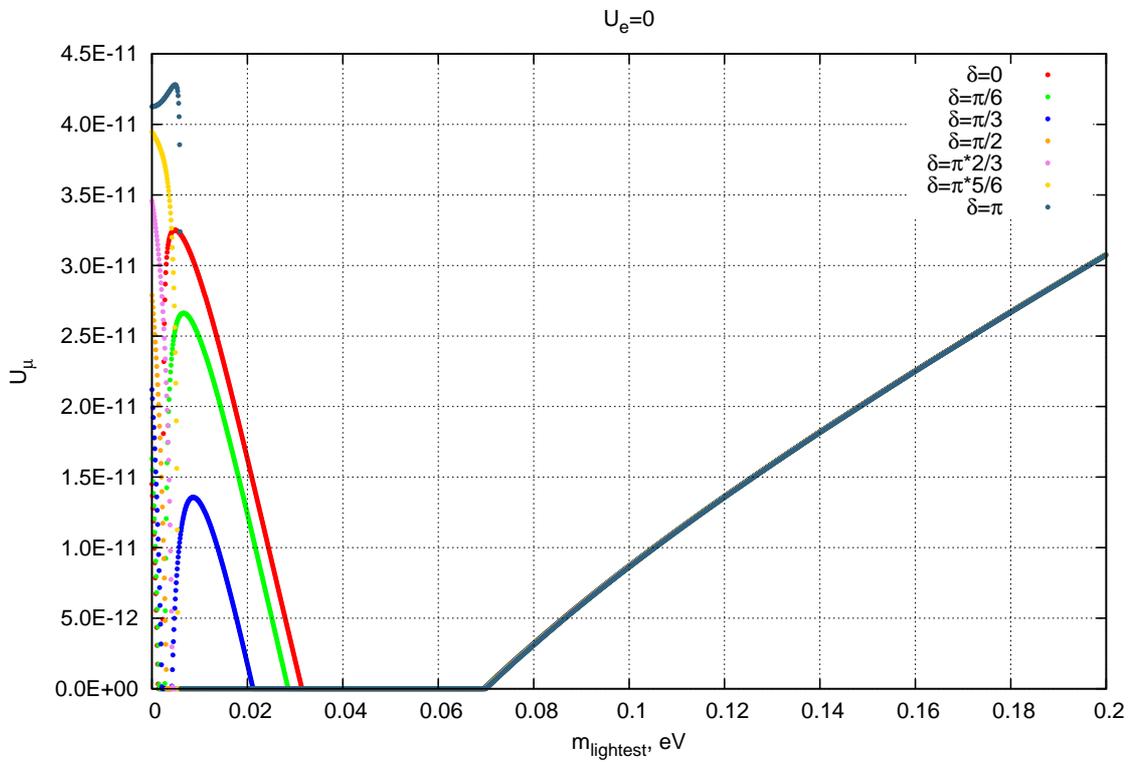}
\caption{Dependence of minimal $U_\mu$ on $m_{lightest}$ at $U_e=0$ for the normal hierarchy. Different curves correspond to different $\delta$ values.}
\end{figure*}

\begin{figure*}
\label{fig13}
\includegraphics[height=0.45\textheight]{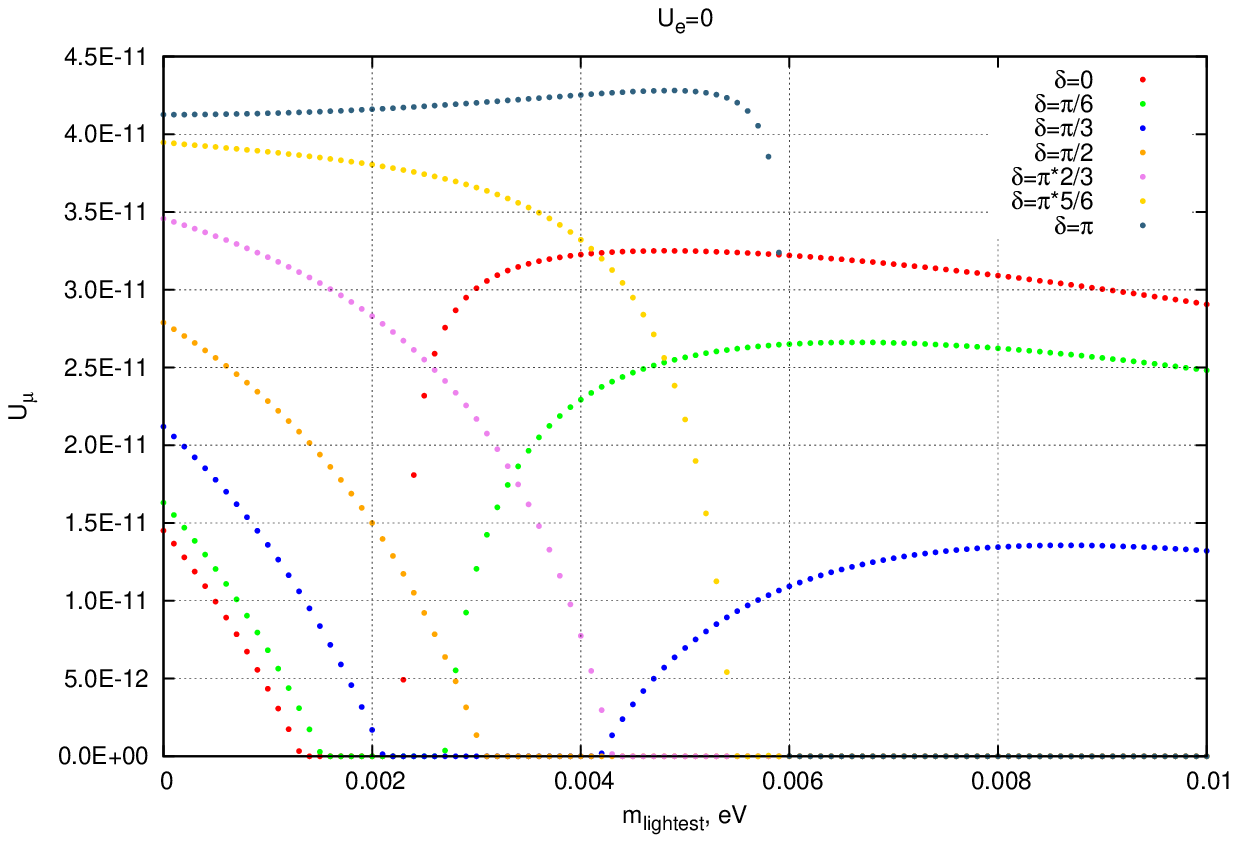}
\caption{Zoom in the small scale area of Fig \ref{fig13}.}
\label{fig13a}
\end{figure*}

\begin{figure*}
\includegraphics[height=0.45\textheight]{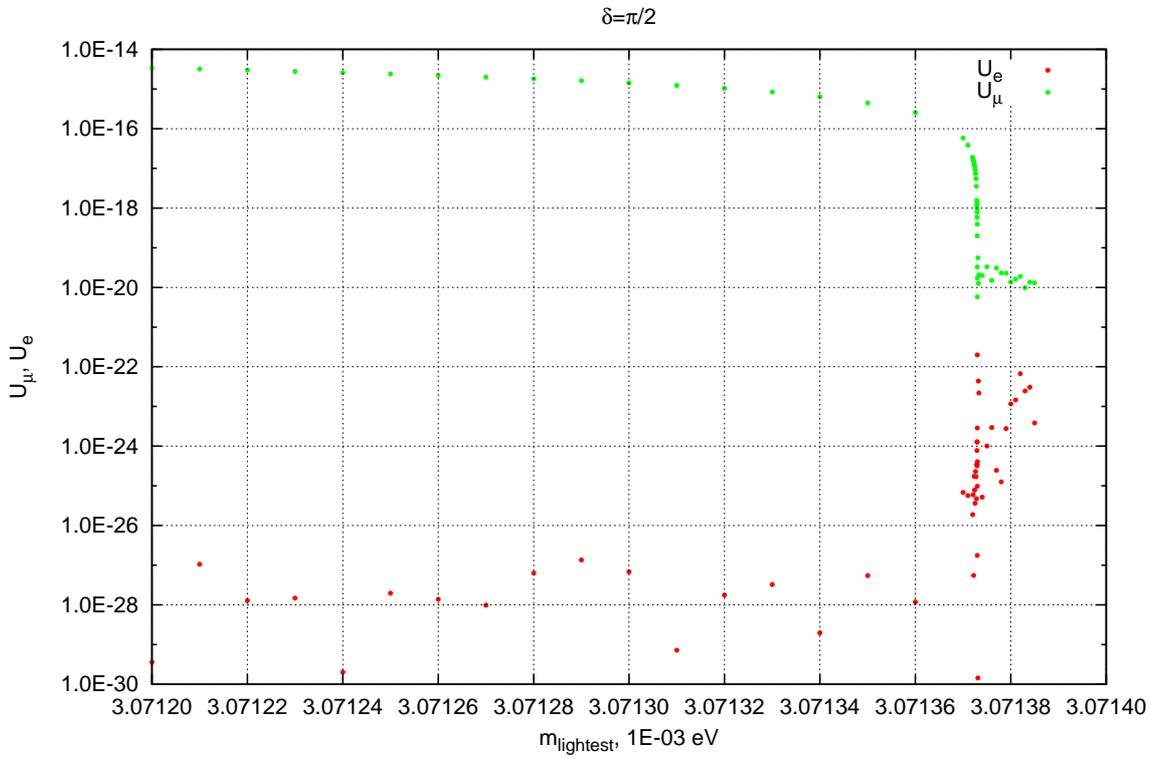}
\caption{Dependence of minimal $U_\mu$ on $m_{lightest}$ at $U_e=0$ in the ``plateau'' proximity for the normal hierarchy. Non-zero values of $U_e$ obtained from formulas from Appendix \ref{zero case}.}
\end{figure*}

\begin{figure*}
\label{fig14}
\includegraphics[height=0.45\textheight]{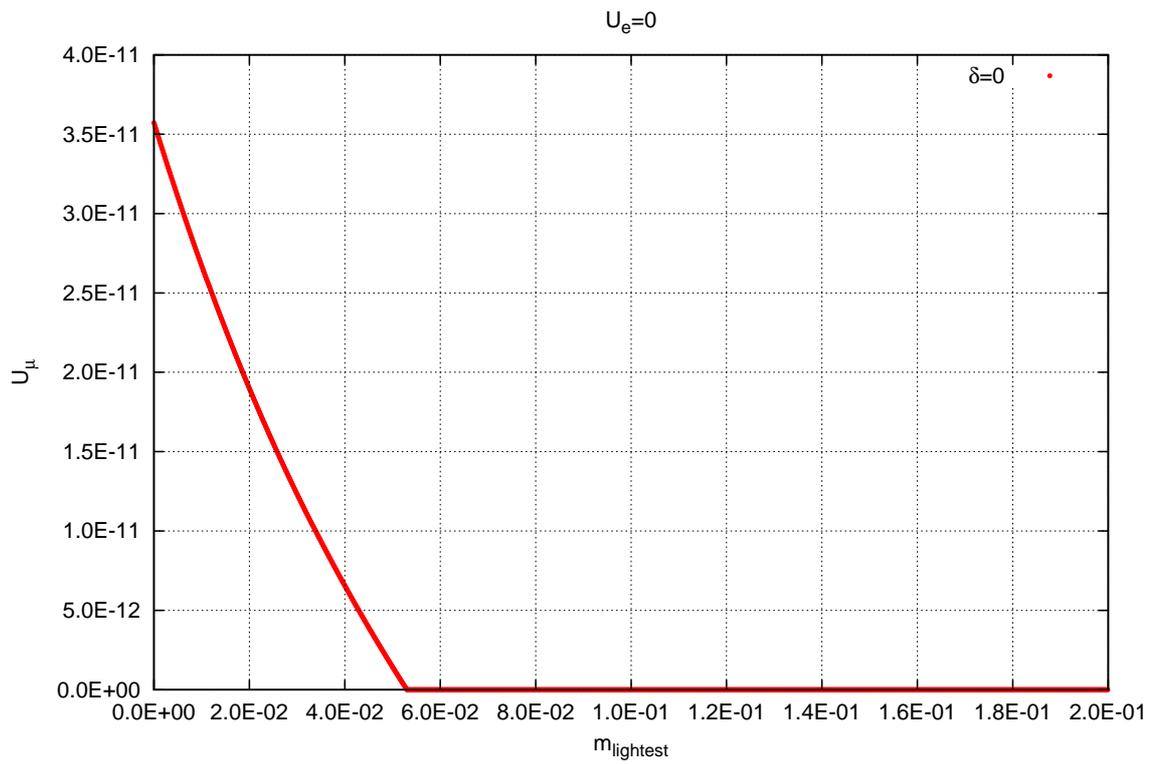}
\caption{Dependence of minimal $U_\mu$ on $m_{lightest}$ at $U_e=0$ and $\delta=0$ for the inverted hierarchy.}
\label{figExtra}
\end{figure*}

\begin{figure*}
\includegraphics[height=0.45\textheight]{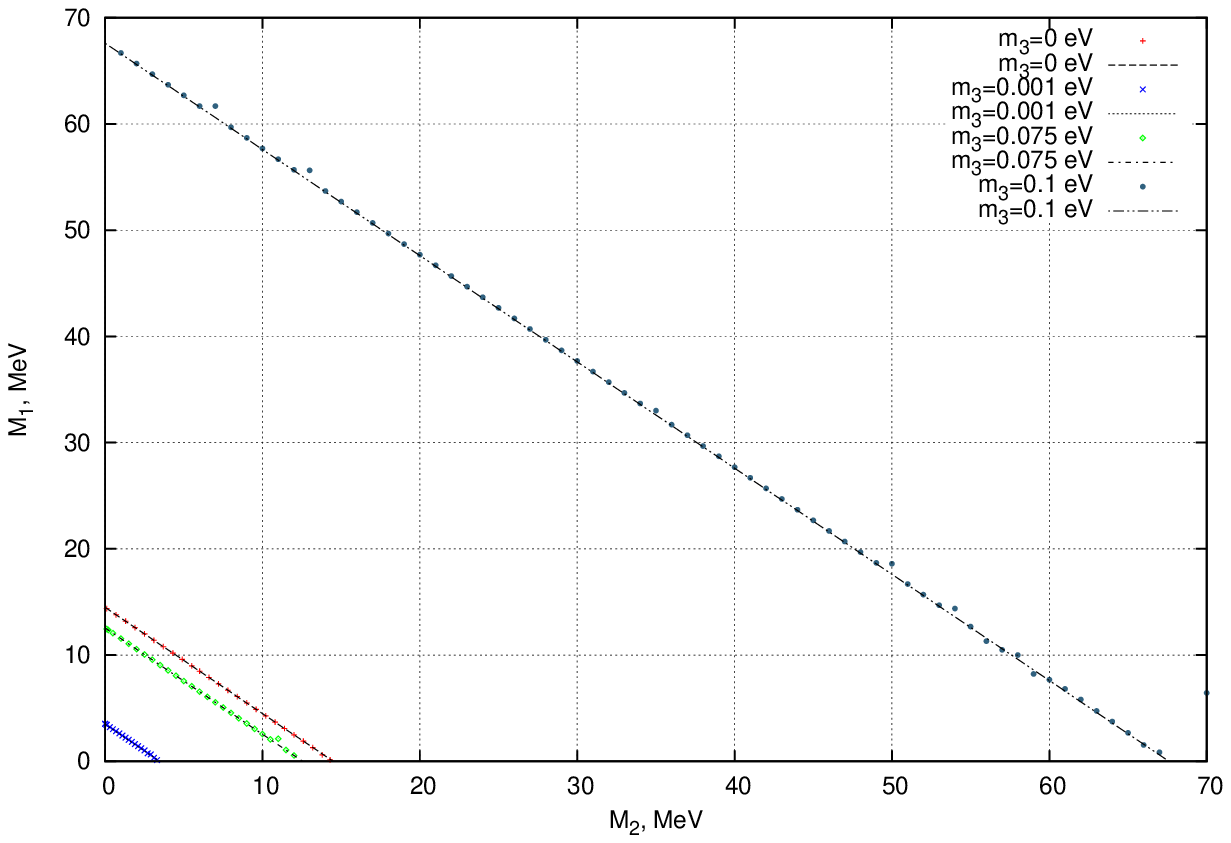}
\caption{($M_1,\,M_2$) space. The regions below the lines will be ruled out by experiments with sensitivity $U_c=5 \times 10^{-11}$ for the normal hierarchy. The different lines correspond to different values of $m_{lightest}$.}
\end{figure*}

\begin{figure*}
\label{fig15}
\includegraphics[height=0.45\textheight]{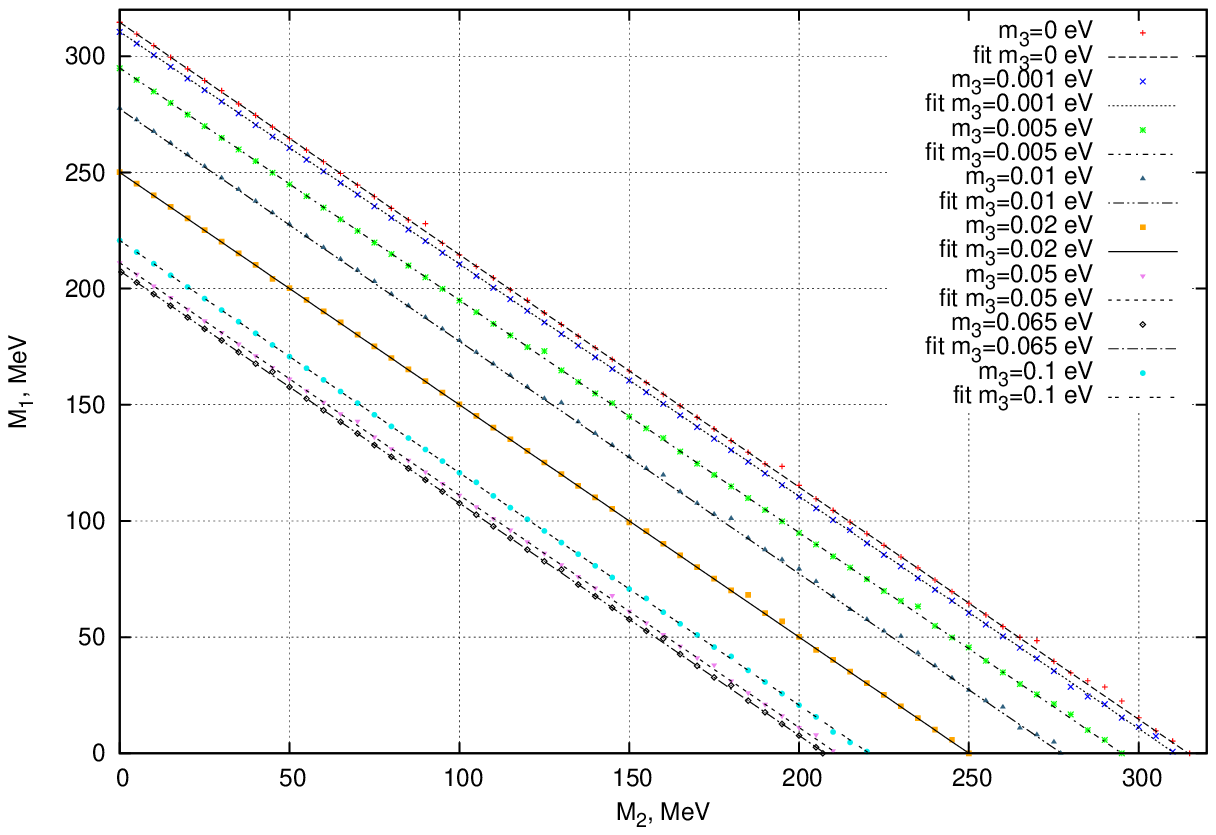}
\caption{($M_1,\,M_2$) space. The regions below the lines will be ruled out by experiments with sensitivity $U_c=5 \times 10^{-11}$ for the inverted hierarchy. The different lines correspond to different values of $m_{lightest}$.}
\label{fig16}
\end{figure*}

\begin{figure*}
\includegraphics[height=0.45\textheight]{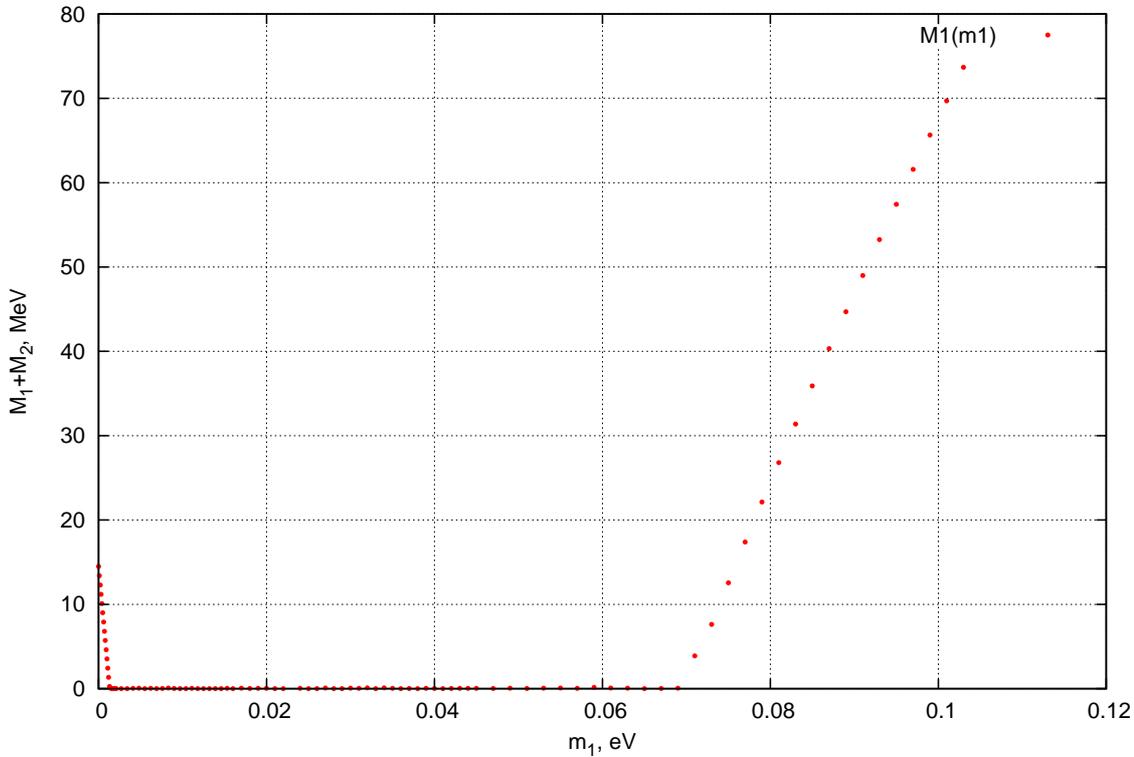}
\caption{Dependence of $M_1+M_2$ on $m_{lightest}$ for the normal hierarchy and $U_c=5 \times 10^{-11}$.}
\label{fig17}
\end{figure*}

\begin{figure*}
\includegraphics[height=0.45\textheight]{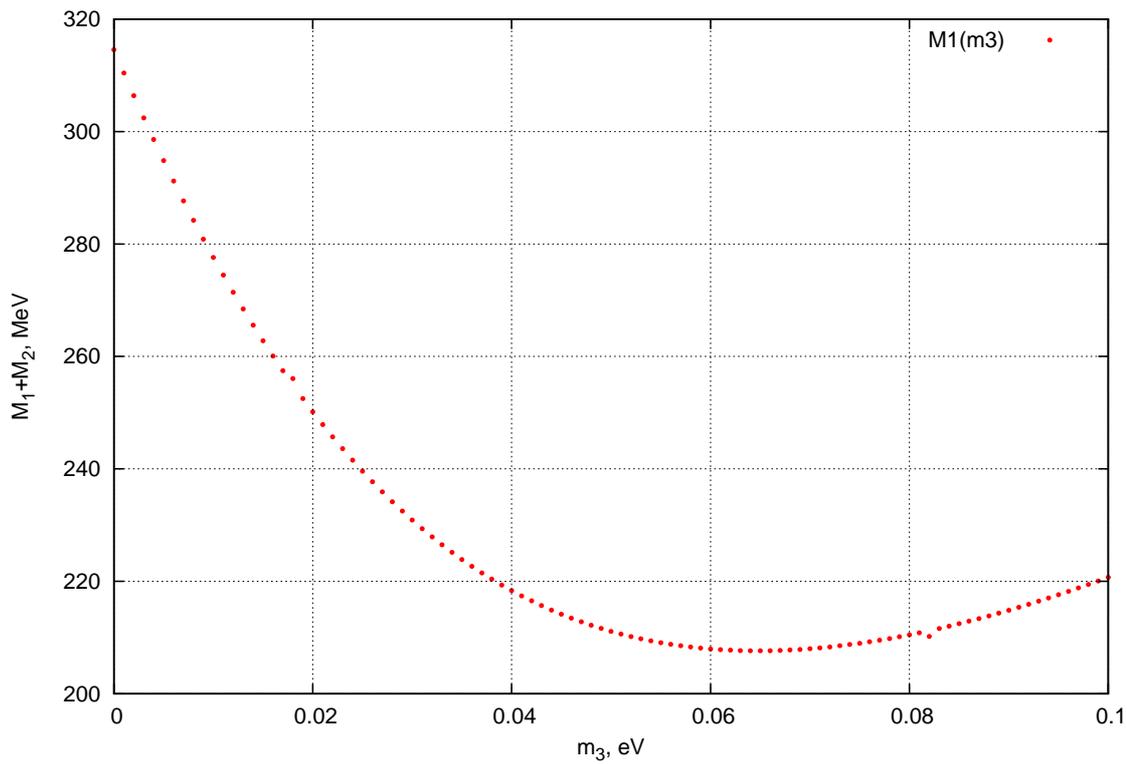}
\caption{Dependence of $M_1+M_2$ on $m_{lightest}$ for the inverted hierarchy and $U_c=5 \times 10^{-11}$.}
\label{fig18}
\end{figure*}

\begin{figure*}
\includegraphics[height=0.45\textheight]{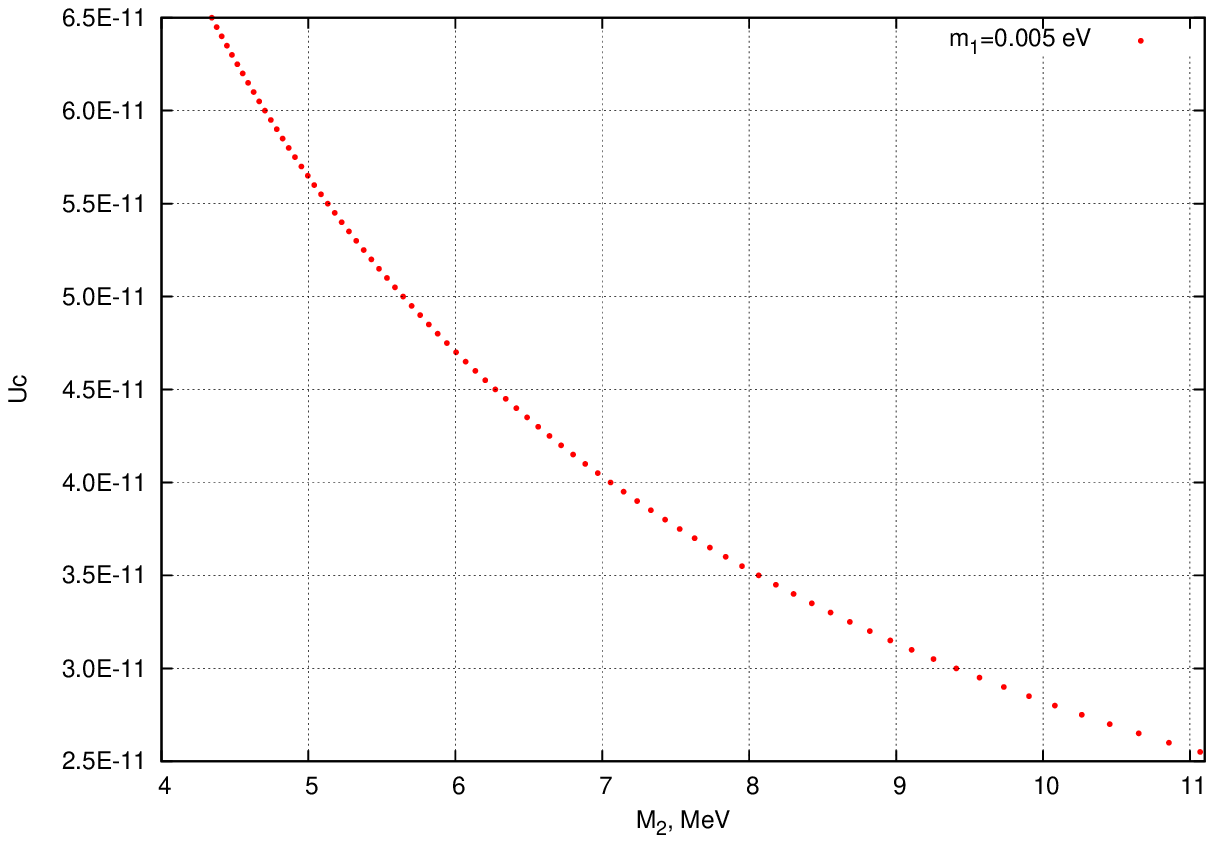}
\caption{Dependence of parameter $U_c$ on the heavier sterile neutrino mass for the normal hierarchy for fixed $M_1=400\,\textrm{MeV}$, $m_{lightest}=0.005\,\textrm{eV}$.}
\end{figure*}

\begin{figure*}
\label{fig19}
\includegraphics[height=0.45\textheight]{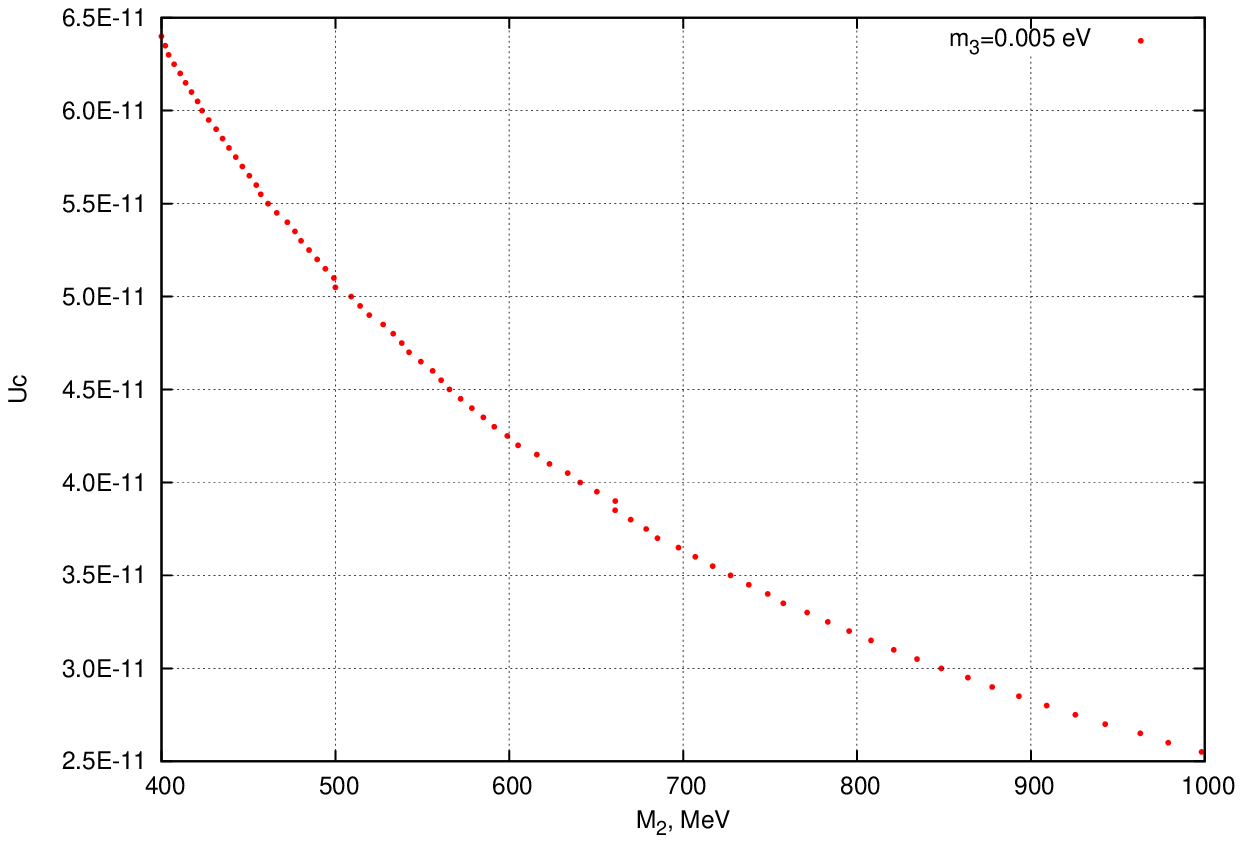}
\caption{Dependence of parameter $U_c$ on the heavier sterile neutrino mass for the inverted hierarchy for fixed $M_1=400\,\textrm{MeV}$, $m_{lightest}=0.005\,\textrm{eV}$.}
\label{fig20}
\end{figure*}

For our numerical simulations we take the observable parameters $\theta_{12}$, $\theta_{13}$, $\theta_{23}$, $\Delta m_{21}^2$, $|\Delta m^2|$ from experimental data analysis \eqref{data2}, \eqref{data1}, and not yet observed quantities $\delta, m_{lightest},  M_1, M_2$ as external parameters.
After specifying these parameters we minimize functions $U_\mu$ and $U_e$ using downhill simplex method taking $z_1, z_2, z_3, \alpha_1, \alpha_2$ as variables.

After obtaining values of parameters which correspond to the minimum, we write them in a file, and restore the value of $U_e, U_\mu$ using eq. \eqref{eq6}, \eqref{eq7}.
Such approach guarantees that our results correspond to the actually possible vales of $U_e$ and $U_\mu$.
Experimentally the mixing $U_e$, $U_\mu$ can be constrained from heavy meson decays as we discussed in Sec. \ref{matrix}, and we obtain the minimal mixing, which is still consistent with the seesaw mechanism.

Concrete formulas used for calculation are laid out in Appendix \ref{app1}.
Note that possible values $\delta \in [0,2 \pi)$ can be restricted to $\delta \in [0, \pi)$, as \eqref{eq7} can't distinguish between some points of this area, as also shown in Appendix \ref{app1}.

To make it easier for applications, in Appendix \ref{fit} we provide a semianalytical approximation for curves presented in Figs. \ref{fig1} -- \ref{fig12a}.
For calculated parameters of these fits see Appendix \ref{fit}.

\subsection{Zero phases}
\label{zero}

Firstly we study the dependence of minimal $U_\mu$ on $U_e$ with zero CP-violating phases for a set of values of $m_{lightest}$ and two hierarchies of masses.
This work was done in \cite{Gorbunov:2013dta}.
To make comparison possible, firstly we perform calculations using parameters, presented in \cite{Gorbunov:2013dta}.
Obtained graphs should have been identical to those of Ref. \cite{Gorbunov:2013dta}, but differences between them have been significant enough to require an explanation.

Through discussion with authors of \cite{Gorbunov:2013dta} we came to a conclusion, that declared values of experimentally observed variables didn't match ones used in calculation  (the adopted values were provided by authors of \cite{Gorbunov:2013dta}):
\begin{displaymath}
\begin{array}{ccc}
|\Delta m^2_{32}|= \left( 5.00\times10^{-2} \textrm{eV} \right)^2\\
\Delta m^2_{21}= \left( 8.75\times10^{-3} \textrm{eV} \right)^2\\
\theta_{12}=33.80^{\circ}\\
\theta_{23}=45.5^{\circ}\\
\theta_{13}=8.82^{\circ}\\
\end{array}
\end{displaymath}
Our graphs, constructed with these parameters, perfectly match ones from Ref. \cite{Gorbunov:2013dta}.

In Figs. \ref{fig7} -- \ref{fig1} we present the results calculated with presently accepted central values of neutrino parameters \eqref{data2}, \eqref{data1} for both hierarchies and we use only these values from now on as well.

We plot dependence of minimal $U_\mu$ on minimal $U_e$ for the normal hierarchy and $\delta=\alpha_1=\alpha_2=0$ and different values of $m_{lightest}$ on Fig. \ref{fig7}.
Fig. \ref{fig7a} is a zoom in a small scale area of Fig. \ref{fig7}, studied in \cite{Gorbunov:2013dta}.
On Fig. \ref{fig1} we plot the same dependence for the inverted hierarchy.

To make further descriptions more tangible, we take for each specific curve  the value of $U_\mu$ when $U_e=0$ and the value of $U_e$ when $U_\mu=0$ and call them characteristic values of $U_\mu$ and $U_e$  for this curve correspondingly.
From the form of the graphs one can see that they represent the maximal values of $U_\mu$ and $U_e$.
Usually we present only the smallest characteristic values.
Thus for the normal hierarchy defined in that way the characteristic values of the seesaw mixing are $U_\mu \approx 1.5 \times 10^{-11}$, $U_e \approx 2 \times 10^{-12}$.
For the inverted hierarchy they are $U_\mu \approx 3.5 \times 10^{-11}$, $U_e \approx 7.5 \times 10^{-11}$.

Basically, curves have the behaviour of ``the greater the mass the higher the curves lay''.
One can see that, for zero CP-violating phases, ``$m_{lightest}=0$'' curve corresponds to the lower limit of the values of mixing.
To fully explore type I seesaw model with corresponding sterile neutrino masses for zero CP-violating phases one just has to reach the sensitivity corresponding to that lower limit.

\subsection{Non-zero phases}
\label{non-zero}

In this section we study the dependence of minimal $U_\mu$ on $U_e$, but with non-zero phases.
We only lay out here some of the more characteristic graphs.

\subsubsection{Inverted hierarchy}
\label{inverted}
We plot dependence of minimal $U_\mu$ on minimal $U_e$ for the inverted hierarchy and different values of $m_{lightest}$ on Fig. \ref{fig6}.
The difference between graphs calculated using different values of $\delta$ can't be observed with naked eye.
As dependence on $\delta$ doesn't play much role for these graphs, we only include graph for minimization on $\delta, \alpha_1, \alpha_2$.

For the inverted hierarchy and minimization on $\delta$, $\alpha_1$, $\alpha_2$ (Fig. \ref{fig6}) a significant difference can be seen as compared with zero phases case (Fig. \ref{fig1}).
First of all while the curve corresponding to $m_{lightest}=0$ practically doesn't change its position, other curves change their behaviour of ``the greater the mass the higher the curves lay'' to the opposite one of ``the greater the mass the lower the curves lay''.
In this way, in case of non-zero CP-violating phases the lower limit corresponds to the curve with the highest possible mass.
From graphs in Fig. \ref{fig6} one can see that with growth of $m_{lightest}$
characteristic values of $U_\mu$, $U_e$ lose as much as several orders of magnitude.
Here we define the characteristic values in the same way as we did it in \ref{zero}.
$m_{lightest}=0$ curve keeps these values at $U_\mu \approx 3.5 \times 10^{-11}$, $U_e \approx 7.5 \times 10^{-11}$, not differing much from $\delta = \alpha_1 = \alpha_2 = 0$ case.
For ``$m_{lightest}=0.05\,$eV'' curve characteristic values are estimated to be $U_\mu \approx 1.4 \times 10^{-12}, U_e \approx 2.7 \times 10^{-11}$.
Characteristic values diminish even more rapidly with further growth of $m_{lightest}$, and for ``$m_{lightest}=0.1\,$eV'' curve these values are indistinguishable from the point of origin,  $U_\mu \sim  U_e \sim 10^{-20}$.
This behaviour exceeds our expectation  and we study dependence on the $m_{lightest}$ in more detail in Sec. \ref{zero U_e}.
Nevertheless, this result shows that the estimation of the upper limit of $m_{lightest}$ can become the leading factor in determining the theoretical lower limit on the mixing angles.
Non-zero values of minimized $\alpha_1, \alpha_2$ are responsible for the difference with the results of Sec. \ref{zero}.

\subsubsection{Normal hierarchy}
\label{normal}
For the normal hierarchy the difference with zero phases case takes more complex shape.

We plot dependence of minimal $U_\mu$ on minimal $U_e$ for the normal hierarchy and different values of $m_{lightest}$ for $\delta = 0, \frac{\pi}{2}, \pi, \delta=1.35 \pi, $ on Figs. \ref{fig8a}, \ref{fig9a}, \ref{fig10a}, \ref{fig11a} respectfully.
$\alpha_1, \alpha_2$ are minimization variables.
On Fig. \ref{fig12a} $\delta$ is also a minimization variable.

For\, zero\, phases\, (Figs.\, \ref{fig7}, \ref{fig7a})\, we\, have\, the ``the greater the mass the higher the curves lay'' behaviour.
It changes completely, as curves start to cross each other, behave differently in areas with big values of $m_{lightest}$ compared to the areas of small values.
For some masses the mixing angle takes such minuscule  values, that the corresponding curves become indistinguishable from the point of origin (they lose several orders of magnitude up to $U_\mu \sim U_e \sim 10^{-20}$). 
Moreover, for different values of $\delta$ the graphs differ significantly from each other.

We study more closely the dependence of graphs on $m_{lightest}$ and $\delta$ to understand such behaviour in the next Sec. \ref{zero U_e}.

\subsubsection{Dependence on $m_{lightest}$}
\label{zero U_e}

As each curve in Figs. \ref{fig8a} -- \ref{fig12a} monotonously declines with growth of $U_e$, for convenience we choose the case of $U_e=0$, representing the highest value of $U_\mu$, as the characteristic point for each curve and study the dependence of $U_\mu$ on $m_{lightest}$.

Firstly,\, we\, plot\, dependence\, of\, minimal\, $U_\mu$ on $m_{lightest}$ for the normal mass hierarchy and $U_e$ set to zero for a set of $\delta$-phases (Figs. \ref{fig13}, \ref{fig13a}).
Starting from the $m_{lightest}=0$ point on the graph, where values of all curves lay at the same magnitude of $10^{-11}$, the curves start to decline as $m_{lightest}$ grows.
The only exception is $\delta=\pi$ curve, which exhibits a short growth before it reaches a local maximum and starts to decline like every other curve.
At this point all curves on this plot show a common behaviour: they decline swiftly, their values lose several orders of magnitude over minuscule increase in $m_{lightest}$. 
For linear scale it seems as if values of $U_\mu$ swiftly decline to zero values\footnote{We should note, that $U_\mu$ and $U_e$ can't reach zero values simultaneously.
It's rather obvious, as one can't obtain matrix with three eigenvalues (active neutrino mass matrix) while using rotational matrix with only two non-zero eigenvalues.} and stay this way for a wide range of values of $m_{lightest}$, before they start to increase just as swiftly as they have declined earlier.
Due to its distinguished form we call this problematic area the ``plateau''.
Another point of interest is that the curves no longer depend on delta after $m_{lightest}$ becomes great enough to exit ``plateau'' area, uniting into one curve, as can be seen on Fig. \ref{fig13}.

We study the ``plateau'' proximity area more closely to understand what values our function can actually reach in the ``plateau''.
We plot dependence of minimal $U_\mu$ on $m_{lightest}$ at $U_e=0$ in the ``plateau'' proximity for the normal hierarchy in logarithmic scale for $\delta=\frac{\pi}{2}$ on Fig. \ref{fig14}.
We obtain all active and sterile neutrino parameters listed in \ref{active}, \ref{sterile} and by putting their numerical values in \eqref{eq6}, we obtain our resulting values for $U_e, U_\mu$ in accordance with formulas in Appendix \ref{zero case}.
Even if we set $U_e=0$ analytically, numerically it can be preserved only to a certain degree.
After reaching the aforementioned values of $U_\mu \sim 10^{-20}$ for $U_\mu$ with $U_e=0$ our numerical calculation has reached the limit of it's accuracy.
It is visible that the closer we get to ``plateau'', the closer is the reconstructed value of ``zero'' $U_e$ to the value of $U_\mu$ itself.
Studying this area of parameters any deeper requires more refined procedure and is beyond the scope of this paper given the fact that mixing $|U_{I\alpha}|^2 \sim 10^{-20}$ can't be tested by experiments in the foreseeable future.

The reason why a small increment in the values of $m_{lightest}$ brings such drastic drop in the values of $U_e, U_\mu$ turns out to be a mutual subtraction.
As we minimize $\alpha_1, \alpha_2$ in plateau area they can be chosen in such a way that $U_{\mu1}, U_{\mu2}$ can loose their leading orders, their absolute values dropping drastically.
On a side note, the mixing with third sterile neutrino in such case can be more intense than mixing with other two, but it won't be observable in considered experiments if $M_3>2$ GeV.

Although we say that we have reached the limit of our calculation's accuracy, it doesn't mean that any of results presented here are inaccurate.
What we mean is that the values are no higher than the ones we provide, but in the ``plateau'' area they can be even lower than $U_\mu \sim 10^{-20}$.

We plot dependence of minimal $U_\mu$ on $m_{lightest}$ for the inverted mass hierarchy and $U_e$ set to zero for $\delta=0$ on Fig. \ref{figExtra}.
For inverted hierarchy there is no significant dependence on $\delta$ for all values of $m_{lightest}$ and so we don't include graphs with other values of $\delta$.
Starting from the $m_{lightest}=0$ point on the graph, where values of the curve lay at the magnitude of $10^{-11}$, the curves start to decline as $m_{lightest}$ grows.
If one looks in linear scale, curve simply reaches ``zero'' value and after that stays in the ``plateau''.
At $m_{lightest}=0.1\,$eV curve still doesn't leave ``plateau''.

\subsection{Dependence on sterile neutrinos masses}
\label{masses}

In this Section we study the dependence of the minimal mixing on the values of sterile neutrino masses.
First of all we notice that explicit dependence of mixing values $U_{\alpha i}$ can be taken from \eqref{eq6}: $U_{\alpha i} \sim \frac{1}{\sqrt M_i}$.
Therefore for degenerative case $M_1=M_2=M$ we simply have $U_e \sim \frac{1}{M},\, U_\mu \sim \frac{1}{M}$ (as already stated in \cite{Gorbunov:2013dta}).

In general case $M_1 \ne M_2$ and mixing may be distributed\, between\, sterile\, neutrinos\, in\, uneven\, manner.
In fact our numerical estimation shows that, for the minimal values of mixing we are interested in, the lighter sterile neutrino is practically decoupled.
We plot ($M_1,\,M_2$) space for the normal hierarchy on Fig. \ref{fig15}.
Likewise, we plot ($M_1,\,M_2$) space for the inverted hierarchy on Fig. \ref{fig16}.
The regions below the lines will be ruled out by experiments with sensitivity $U_c=5 \times 10^{-11}$ for the normal hierarchy. The different lines correspond to different values of $m_{lightest}$.
Parameter $U_c$ represents the sensitivity of the experiment needed to rule out the seesaw model in a specified sterile neutrino mass region.
From these Figs. one can see that the lines corresponding to the minimal possible values of $M_1,\,M_2$ follow the equation $M_1+M_2 = f(m_{lightest})/U_c$.
This behaviour is the same as one found in  \cite{Gorbunov:2013dta}, although concrete dependence is modified with the introduction of parameters $\delta, \alpha_1, \alpha_2$.

For the normal hierarchy one can see that lines start to lay lower with the growth of $m_{lightest}$ until they reach zero in ``plateau'' area, and start to lay higher with the growth of $m_{lightest}$ after $m_{lightest}$ exits plateau values.
For the inverted hierarchy lines lay lower with the growth of $m_{lightest}$ until they reach the minimum value at $m_{lightest}=0.065\,$eV and start to lay higher with the growth of $m_{lightest}$ after that.

We plot dependence of $M_1+M_2$ on $m_{lightest}$ for $U_c=5 \times 10^{-11}$ for the normal hierarchy on Fig. \ref{fig17} and inverted hierarchy on Fig. \ref{fig18}.
The mentioned above behaviour can be seen on this graphs in more detail.

We plot dependence of parameter $U_c$ on the heavier sterile neutrino mass for the normal hierarchy needed to rule out seesaw model for fixed $M_1=400\,\textrm{MeV}$, $m_{lightest}=0.005\,\textrm{eV}$.
Dependence of parameter $U_c$ on the heavier sterile neutrino mass for the inverted hierarchy for fixed $M_1=400\,\textrm{MeV}$, $m_{lightest}=0.005\,\textrm{eV}$.
This way $M_2$ is the heavier mass in region $400\,\textrm{MeV} < M_2 < 1\, \textrm{GeV}$.
One can see that the minimal value of $U_c$ monotonically decreases with the growth of the heavier sterile neutrino mass.
Minimal mixing values practically don't depend on the lighter sterile neutrino mass.

\section{Conclusion}
In this paper we study minimal possible mixing angles between sterile and active neutrinos $|U_{I\alpha}|^2$ for the specific case of two sterile neutrinos with masses less than 2 GeV.
These angles provide us with information on sensitivity which experiments such as SHiP or DUNE or their successors should achieve to fully explore type I seesaw model with two sterile neutrinos with masses below 2 GeV and one undetectable sterile neutrino.
To that end we study the dependence of mixing matrix on model parameters ($\delta, \alpha_1, \alpha_2$), that hasn't been considered in work \cite{Gorbunov:2013dta}.
Characteristic values for zero phases are $|U_{I\alpha}|^2 \sim 10^{-11}$.
Introducing the dependence on CP-violating phases, we observe strong dependence on the lightest neutrino mass $m_{lightest}$ and these phases.
For both hierarchies minimal mixing $|U_{I\alpha}|^2$ could be lowered depending on $m_{lightest}$ and ($\delta, \alpha_1, \alpha_2$) to the values of $10^{-20}$ at least.
These results can be rescaled to other values of sterile neutrinos masses: if we simultaneously change $M_I \to X M_I$ (for all three  sterile neutrinos), than mixing also simply changes by that factor: $U_e \to \frac{1}{X} U_e, U_\mu \to \frac{1}{X} U_\mu$.
Such sterile neutrinos can be produced in the decays of weak gauge bosons and other heavy SM particles, e.g. in LHC, FCC.
We conclude that still unknown parameters of active neutrino $m_{lightest}, \delta, \alpha_1, \alpha_2$ may significantly change the mixing pattern and should be taken into account in future experiments.


\begin{acknowledgements}
We would like to thank D. Gorbunov and A. Panin for valuable discussions.
The work was supported by the RSF grant 17-12-01547.
\end{acknowledgements}

\appendix
\section{Formulas used for calculation}
\label{app1}
If one takes formulas (\ref{eq3}), (\ref{eq5}), (\ref{eq6}) and writes down (\ref{eq7}) using them, one can obtain the following equations:
\begin{eqnarray}
\label{eq8}
U_e & = & \frac{1}{M_1} |\lambda_1 c_2 + \lambda_2 s_2|^2 + \nonumber\\
&&+ \frac{1}{M_2} |\lambda_3 c_3 + (\lambda_2 c_2-\lambda_1 s_2) s_3|^2\\
\label{eq9}
U_\mu & = & \frac{1}{M_1} |\eta_1 c_2 + \eta_2 s_2|^2 +\nonumber\\
&&+ \frac{1}{M_2} |\eta_3 c_3 + (\eta_2 c_2-\eta_1 s_2) s_3|^2,
\end{eqnarray}
where:
\begin{eqnarray}
\label{eq10}
\lambda_1 & = &  \sqrt{m_1} A_{11} c_1 + \sqrt{m_2} A_{21} s_1\\
\label{eq11}
\lambda_2 & = &  \sqrt{m_3} A_{31}\\
\label{eq12}
\lambda_3 & = & -\sqrt{m_1} A_{11} s_1 + \sqrt{m_2} A_{21} c_1\\ 
\label{eq13}
\eta_1    & = &  \sqrt{m_1} A_{12} c_1 + \sqrt{m_2} A_{22} s_1\\
\label{eq14}
\eta_2    & = &  \sqrt{m_3} A_{32}\\
\label{eq15}
\eta_3    & = & -\sqrt{m_1} A_{12} s_1 + \sqrt{m_2} A_{22} c_1
\end{eqnarray}
Here for simplicity we introduced $A=U_{PMNS}^\dagger$.

One can notice that the following reflections don't change the values of $U_\mu$, $U_e$:
\begin{equation}
\label{eq40}
\left\{\begin{array}{ccl}
\delta & \to & -\delta\\
\alpha_1 & \to & -\alpha_1\\
\alpha_2 & \to & -\alpha_2\\
z_1 & \to & z_1^*\\
z_2 & \to & z_2^*\\
z_3 & \to & z_3^*
\end{array}\right. \Rightarrow \left\{\begin{array}{l}
U_\mu \to U_\mu\\
U_e \to U_e
\end{array}\right..
\end{equation}
Hence, in order to explore all the possible values of $U_e, U_\mu$ one can restrict possible values of $\delta$ as follows: $\delta \in [0, \pi)$.

\subsection{Direct expression of $U_\mu$ dependence on $U_e$}
\label{app2}
If we want to find dependence of minimal value of $U_\mu$ on $U_e$, we can simply solve equation (\ref{eq8}) for $\mathrm{Im}[z_3]$.
First we write it down in the following way:
\begin{eqnarray}
\label{eq16}
U_e & = & f_1 \sinh^2 y_3 - 2 f_2 \sinh y_3\, \cosh y_3 + f_3\\
\label{eq17}
U_\mu & = & f_4 \sinh^2 y_3 - 2 f_5 \sinh y_3\, \cosh y_3 + f_6,
\end{eqnarray}
where $z_j= x_j+i y_j,\, j=1,2,3$ and:
\begin{eqnarray}
\label{eq18}
f_1 & = & \frac{1}{M_2} \Big(|\lambda_3|^2 + |\lambda_2 c_2 - \lambda_1 s_2|^2 \Big)\\
\label{eq19}
f_2 & = & \frac{1}{M_2} \mathrm{Im}[\lambda_3^*(\lambda_2 c_2 - \lambda_1 s_2)]\\
\label{eq20}
f_3 & = & \frac{1}{M_2} \Big(|\lambda_3|^2 \cos^2 x_3 + |\lambda_2 c_2 - \lambda_1 s_2|^2 \sin^2 x_3 + {}
\nonumber\\
& & {} + 2 \sin x_3 \cos x_3 \mathrm{Re}[\lambda_3^*(\lambda_2 c_2 - \lambda_1 s_2)] \Big) + {}
\nonumber\\
& & {} + \frac{1}{M_1} |\lambda_1 c_2 + \lambda_2 s_2|^2\\
\label{eq21}
f_4 & = & \frac{1}{M_2} \Big( |\eta_3|^2 + |\eta_2 c_2 - \eta_1 s_2|^2 \Big)\\
\label{eq22}
f_5 & = & \frac{1}{M_2} \mathrm{Im}[\eta_3^*(\eta_2 c_2 - \eta_1 s_2)]\\
\label{eq23}
f_6 & = & \frac{1}{M_2} \Big(|\eta_3|^2 \cos^2 x_3 + |\eta_2 c_2 - \eta_1 s_2|^2 \sin^2 x_3 + {}
\nonumber\\
& & {} + 2 \sin x_3  \cos x_3  \mathrm{Re}[\eta_3^*(\eta_2 c_2 - \eta_1 s_2)] \Big) + {}
\nonumber\\
& & {} + \frac{1}{M_1} |\eta_1 c_2 + \eta_2 s_2|^2
\end{eqnarray}

We can change (\ref{eq16}) into a quadratic equation on $\tanh y_3$:
\begin{equation}
\label{eq24}
(f_1-f_3+U_e) \tanh^2 y_3 -2 f_2 \tanh y_3 +f_3-U_e = 0
\end{equation}

And solve it:
\begin{eqnarray}
\label{eq25}
\tanh y_3 & = & \frac{f_2 \pm \sqrt{f_2^2 + (f_1 - f_3 + U_e) (U_e-f_3)}}{f_1-f_3+U_e}\\
\label{eq26}
y_3 & = & \frac{1}{2} \mathrm{ln}\Big(\frac{1+\tanh y_3}{1-\tanh y_3}\Big)
\end{eqnarray}

Applying formulas (\ref{eq10}) - (\ref{eq15}), (\ref{eq18}) - (\ref{eq23}), (\ref{eq25}) to (\ref{eq26}) we can restore value of $y_3$ from the values of $U_e$ and $x_1,x_2,x_3,y_1,y_2,\alpha_1, \alpha_2$.
As we minimize function over these parameters, we can simply exclude all regions, that can't satisfy equation (\ref{eq16}) for the value of $U_e$ we are interested in and minimize (\ref{eq17}) using $y_3$ as a function composition.

One can notice, that simply by replacing all $\lambda_i \leftrightarrow \eta_i$ we can switch from dependence of $U_\mu$ on $U_e$ to the dependence of $U_e$ on $U_\mu$.

\subsection{$U_e=0$ or $U_\mu=0$ case}
\label{zero case}
We mention in Sec. \ref{normal} that minimal $U_\mu$ monotonous\-ly declines with the growth of minimal $U_e$ (see Figs. \ref{fig1} -- \ref{fig12a}).
Therefore it is convenient to use values of $U_\mu$ at $U_e=0$ and values of $U_e$ at $U_\mu=0$ as the characteristic points for each curve.
This situation is also analytically unique, as $U_e=0$ can be transformed into two rather simple complex equations, in contrast with the usual rather complicated single equation $U_e=U_{e0} \neq 0$.
Obviously the same goes for $U_\mu=0$ case, so, to study that case, one can just change $\lambda_i \leftrightarrow \eta_i$ in the following formulas.

\begin{eqnarray}
\label{eq27}
\lambda_1 c_2 + \lambda_2 s_2 = 0\\
\label{eq28}
\lambda_3 c_3 + (\lambda_2 c_2-\lambda_1 s_2) s_3 = 0
\end{eqnarray}
Thus we can express $z_2, z_3$ in terms of $z_1, \alpha_1, \alpha_2$.

We can transform (\ref{eq9}) using (\ref{eq27}, \ref{eq28}) into a more simple form:
\begin{eqnarray}
\label{eq37}
U_\mu & = & \frac{1}{M_1} \frac{|\eta_1 \lambda_2 - \eta_2 \lambda_1|^2}{|\lambda_1^2+\lambda_2^2|} + \nonumber\\
 &&+ \frac{1}{M_2} \frac{|\eta_3 (\lambda_1^2+\lambda_2^2) - \lambda_3  (\eta_2 \lambda_2 + \eta_1 \lambda_1)|^2}{|\lambda_1^2+\lambda_2^2| |\lambda_1^2+\lambda_2^2+\lambda_3^2|},
\end{eqnarray}

Directly solving (\ref{eq27}) one can express $z_2$ as:
\begin{eqnarray}
\label{eq29}
\tan z_2 & = & - \sqrt{\frac{m_1}{m_3}} \frac{A_{11}}{A_{31}} c_1 - \sqrt{\frac{m_2}{m_3}} \frac{A_{21}}{A_{31}} s_1 \equiv \chi_1\\
\label{eq30}
\psi_1 & \equiv & 2 \frac{\mathrm{Im}[\chi_1]}{1+|\chi_1|^2}\\
\label{eq31}
x_2 & = & \frac{\mathrm{sgn}(\mathrm{Re}[\chi_1])}{2} \mathrm{acos}\Big(\frac{1-|\chi_1|^2}{(1+|\chi_1|^2) \sqrt{1-\psi_1^2}}\Big)\\
\label{eq32}
y_2 & = & \frac{1}{4}\mathrm{ln}\Big(\frac{1+\psi_1}{1-\psi_1}\Big)
\end{eqnarray}

Assuming that we have already determined $z_2$, the expression for $z_3$ can be obtained in the same way by solving (\ref{eq28}):

\begin{eqnarray}
\label{eq33}
\tan z_3 & = & \frac{1}{c_2} \frac{1}{\chi_1^2 + 1 }\Big(\sqrt{\frac{m_1}{m_3}} \frac{A_{11}}{A_{31}} s_1 - \sqrt{\frac{m_2}{m_3}} \frac{A_{21}}{A_{31}} c_1 \Big)\\
&& \chi_2 \equiv \tan z_3 \nonumber\\
\label{eq34}
\psi_2 & \equiv & 2 \frac{\mathrm{Im}[\chi_2]}{1+|\chi_2|^2}\\
\label{eq35}
x_3 & = & \frac{\mathrm{sgn}(\mathrm{Re}[\chi_2])}{2} \mathrm{acos}\Big(\frac{1-|\chi_2|^2}{(1+|\chi_2|^2) \sqrt{1-\psi_2^2}}\Big)\\
\label{eq36}
y_3 & = & \frac{1}{4}\mathrm{ln}\Big(\frac{1+\psi_2}{1-\psi_2}\Big)
\end{eqnarray}

We use these equations to reconstruct $U_e, U_\mu$ by means of definition \eqref{eq8}, \eqref{eq9} and compare the results with 0 and \eqref{eq37}, respectively.

\section{Semianalytic approximation of graphs}
\label{fit}
	For convenience, we approximated the numerical results by the function  $$y=A_{2}+\frac{(A_{1}-A_{2})}{1+(\frac{x}{x_{0}})^{p}}\,.$$ Tables \ref{tab1}, \ref{tab2} below  show the approximation coefficients for each graph in the normal and inverted hierarchies, as well as the coefficient $\chi^{2}/n $ and and the number of independent points $n$.

\renewcommand{\arraystretch}{1.25}
\begin{table*}
\caption{\label{tab1} Normal hierarchy}
\begin{tabular}{|l|*{8}{c|}}
\cline{1-8}	
&$m_{1}$, eV &$A_{1}$ &$A_{2}$ &$x_{0}$ &p & $\chi^{2}/n$&n\\
\cline{1-8}

\multirow{7}{*}{$\delta=\alpha_1=\alpha_2=0$}&$0$ & $1.42\times10^{-11}$ & $-7.34\times10^{-12}$  & $6.86\times10^{-13}$ & 0.67&1.06&162\\
\cline{2-8}
&$1\times10^{-3}$ & $2.1\times10^{-11}$  & $-1.31\times10^{-11}$ & $1.64\times10^{-12}$ & 0.69&1.08&214\\
\cline{2-8}
&$5\times10^{-3}$ & $3.49\times10^{-11}$  & $-3.64\times10^{-11}$ & $9.82\times10^{-12}$ & 0.76&1.02&133\\
\cline{2-8}
&$1\times10^{-2}$ & $4.34\times10^{-11}$  & $-6.61\times10^{-11}$ & $2.83\times10^{-11}$ & 0.85&1.04& 101\\
\cline{2-8}
&$2\times10^{-2}$ & $5.68\times10^{-11}$  & $-2.09\times10^{-10}$ & $1.47\times10^{-10}$ & 0.88&1.06&98\\
\cline{2-8}
&$5\times10^{-2}$ & $9.81\times10^{-11}$  & $-1.35\times10^{-9}$ & $1.29\times10^{-9}$ & 0.96&3.21&109\\
\cline{2-8}
&$1\times10^{-1}$ & $1.75\times10^{-10}$  & $-2.24\times10^{-8}$ & $2.26\times10^{-8}$ & 0.99&1.2&110\\
\cline{1-8}

\multirow{6}{*}{$\delta=0$, minimization in $\alpha_1, \alpha_2$}&$0$ & $1.43\times10^{-11}$ & $-8.48\times10^{-12}$  & $8.2\times10^{-13}$ & 0.64 &1.06& 88\\
\cline{2-8}
&$1\times10^{-3}$ & $4.3\times10^{-12}$  & $-1.68\times10^{-12}$ & $1.1\times10^{-13}$ & 0.67 &1.03 & 72\\
\cline{2-8}
&$5\times10^{-3}$ & $3.23\times10^{-11}$  & $-3.54\times10^{-10}$ & $3.31\times10^{-11}$ & 0.96&5.11& 226\\
\cline{2-8}
&$1\times10^{-2}$ & $2.86\times10^{-11}$  & $-3.9\times10^{-11}$ & $8.45\times10^{-12}$ & 0.76&1.04&82\\
\cline{2-8}
&$2\times10^{-2}$ & $1.6\times10^{-11}$  & $-8.83\times10^{-12}$ & $2.51\times10^{-12}$ & 0.69&1.03&99\\
\cline{2-8}
&$1\times10^{-1}$ & $8.41\times10^{-12}$  & $-3.26\times10^{-12}$ & $2.59\times10^{-11}$ & 0.85&1.02&155\\
\cline{1-8}

\multirow{3}{*}{$\delta=\pi/2$, minimization in $\alpha_1, \alpha_2$}&$0$ & $2.73\times10^{-11}$ & $-2.31\times10^{-11}$  & $2.78\times10^{-12}$ & 0.7&1.02&197\\
\cline{2-8}
&$1\times10^{-3}$ & $2.24\times10^{-11}$  & $-1.56\times10^{-11}$ & $1.38\times10^{-12}$ & 0.68&1.02&146\\
\cline{2-8}
&$1\times10^{-1}$ & $8.57\times10^{-12}$  & $-3.65\times10^{-12}$ & $2.67\times10^{-11}$ & 0.79&1.04&163\\
\cline{1-8}

\multirow{4}{*}{$\delta=\pi$, minimization in $\alpha_1, \alpha_2$}&$0$ & $4.07\times10^{-11}$ & $-1.14\times10^{-10}$  & $1.83\times10^{-11}$ & 0.86&1.07&70\\
\cline{2-8}
&$1\times10^{-3}$ & $4.07\times10^{-11}$  & $-1.1\times10^{-10}$ & $1.4\times10^{-11}$ & 0.89&1.13&65\\
\cline{2-8}
&$5\times10^{-3}$ & $4.22\times10^{-11}$  & $-3.71\times10^{-10}$ & $7.04\times10^{-12}$ & 1.04&1.05&132\\
\cline{2-8}
&$1\times10^{-1}$ & $8.51\times10^{-12}$  & $-5.12\times10^{-12}$ & $3.68\times10^{-11}$ & 0.69&1.03&119\\
\cline{1-8}

\multirow{3}{*}{$\delta=1.35\pi$, minimization in $\alpha_1, \alpha_2$}&$0$ & $3.32\times10^{-11}$ & $-3.26\times10^{-11}$  & $4.27\times10^{-12}$ & 0.77&1.08&51\\
\cline{2-8}
&$1\times10^{-3}$ & $3.07\times10^{-11}$  & $-2.39\times10^{-11}$ & $2.38\times10^{-12}$ & 0.75&1.09&43\\
\cline{2-8}
&$1\times10^{-1}$ & $8.34\times10^{-12}$  & $-3.57\times10^{-12}$ & $2.73\times10^{-11}$ & 0.79&1.03&103\\
\cline{1-8}

\multirow{3}{*}{minimization in $\delta, \alpha_1, \alpha_2$}&$0$ & $1.4\times10^{-11}$ & $-5.84\times10^{-12}$  & $5.69\times10^{-13}$ & 0.74&1.03&88\\
\cline{2-8}
&$1\times10^{-3}$ & $4.26\times10^{-12}$  & $-1.75\times10^{-12}$ & $1.19\times10^{-13}$ & 0.67&1.03& 100\\
\cline{2-8}
&$1\times10^{-1}$ & $8.42\times10^{-12}$  & $-4.92\times10^{-12}$ & $3.68\times10^{-11}$ & 0.71&1.37&95\\
\cline{1-8}
\end{tabular} 
\end{table*}

\begin{table*}
\caption{\label{tab2} Inverted hierarchy}
\begin{tabular}{|l|*{8}{c|}}
\cline{1-8}	
&$m_{1}$, eV &$A_{1}$ &$A_{2}$ &$x_{0}$ &p & $\chi^{2}/n$&n\\
\cline{1-8}

\multirow{7}{*}{$\delta=\alpha_{1}=\alpha_{2}=0$}&$0$ & $3.5\times10^{-11}$ & $-9.42\times10^{-11}$  & $2.26\times10^{-10}$ & 0.9&1.12& 98\\
\cline{2-8}
&$1\times10^{-3}$ & $3.61\times10^{-11}$  & $-1.02\times10^{-10}$ & $2.41\times10^{-10}$ & 0.9&1.16& 98\\
\cline{2-8}
&$5\times10^{-3}$ & $4.01\times10^{-11}$  & $-1.33\times10^{-10}$ & $2.88\times10^{-10}$ & 0.91&1.08&98\\
\cline{2-8}
&$1\times10^{-2}$ & $4.54\times10^{-11}$  & $-1.83\times10^{-10}$ & $3.61\times10^{-10}$ & 0.92&1.07&98\\
\cline{2-8}
&$2\times10^{-2}$ & $5.71\times10^{-11}$  & $-3.38\times10^{-10}$ & $5.68\times10^{-10}$ & 0.94&1.84&98\\
\cline{2-8}
&$5\times10^{-2}$ & $9.79\times10^{-11}$  & $-1.57\times10^{-9}$ & $2\times10^{-9}$ & 0.97&3.23&98\\
\cline{2-8}
&$1\times10^{-1}$ & $1.75\times10^{-10}$  & $-8.73\times10^{-9}$ & $9.66\times10^{-9}$ & 0.99&4.31& 98\\
\cline{1-8}

\multirow{6}{*}{$\delta=1.32\pi$, minimization in $\alpha_1, \alpha_2$}&$0$ & $3.49\times10^{-11}$ & $-8.03\times10^{-11}$  & $1.87\times10^{-10}$ & 0.92&1.77& 96\\
\cline{2-8}
&$1\times10^{-3}$ & $3.38\times10^{-11}$  & $-7.96\times10^{-11}$ & $1.9\times10^{-10}$ & 0.91&1.4&94\\
\cline{2-8}
&$5\times10^{-3}$ & $3.01\times10^{-11}$  & $-5.86\times10^{-11}$ & $1.53\times10^{-10}$ & 0.91&1.09&94\\
\cline{2-8}
&$1\times10^{-2}$ & $2.61\times10^{-11}$  & $-4.92\times10^{-11}$ & $1.53\times10^{-10}$ & 0.86&1.16& 95\\
\cline{2-8}
&$2\times10^{-2}$ & $1.85\times10^{-11}$  & $-2.36\times10^{-11}$ & $9.64\times10^{-11}$ & 0.83&1.03&95\\
\cline{2-8}
&$5\times10^{-2}$ & $1.33\times10^{-12}$  & $-4.36\times10^{-13}$ & $5.54\times10^{-12}$ & 0.77&1.03&96\\
\cline{1-8}
			
\multirow{6}{*}{minimization in $\delta, \alpha_1, \alpha_2$}&$0$ & $3.51\times10^{-11}$ & $-9.45\times10^{-11}$  & $2.29\times10^{-10}$ & 0.88&1.17&95\\
\cline{2-8}
&$1\times10^{-3}$ & $3.42\times10^{-11}$  & $-8.82\times10^{-11}$ & $2.19\times10^{-10}$ & 0.88&1.16&95\\
\cline{2-8}
&$5\times10^{-3}$ & $3.04\times10^{-11}$  & $-6.85\times10^{-11}$ & $1.88\times10^{-10}$ & 0.87&1.23&95\\
\cline{2-8}
&$1\times10^{-2}$ & $2.61\times10^{-11}$  & $-4.83\times10^{-11}$ & $1.5\times10^{-10}$ & 0.855&1.07&95\\
\cline{2-8}
&$2\times10^{-2}$ & $1.84\times10^{-11}$  & $-2.32\times10^{-11}$ & $9.46\times10^{-11}$ & 0.83&1.06&95\\
\cline{2-8}
&$5\times10^{-2}$ & $1.31\times10^{-12}$  & $-4.39\times10^{-13}$ & $5.49\times10^{-12}$ & 0.77&1.05&95\\
\cline{1-8}
\end{tabular}
\end{table*}

\end{document}